\documentclass{scrartcl}
\usepackage{mystyle}
\usepackage{mymath}
\usepackage[top=1.3in, bottom=1.4in, left=1.1in, right=1.1in]{geometry}

\DeclareMathOperator{\hist}{\text{Hist}}
\usepackage{booktabs}

\usepackage{tabularx}
\newcolumntype{C}{>{\centering\arraybackslash}X} % centered version of 'X' columns

\title{\Large\rmfamily Extreme events and their optimal mitigation in nonlinear structural systems excited by stochastic loads: Application to ocean engineering systems}
\author{Han Kyul Joo, Mustafa A. Mohamad, Themistoklis P. Sapsis\thanks{Corresponding
author: \href{mailto:sapsis@mit.edu}{sapsis@mit.edu},
        Tel: (617) 324-7508, Fax: (617) 253-8689%
        }
\texorpdfstring{\vspace{0.25em}\\
Department of Mechanical Engineering, \\
Massachusetts Institute of Technology, \\
77 Massachusetts Ave., Cambridge, MA 02139}{}}
\date{\today}

\begin{document}

\maketitle
\begin{abstract}
We develop an efficient numerical method for the probabilistic quantification of the response statistics of nonlinear multi-degree-of-freedom structural systems under extreme forcing events, emphasizing accurate heavy-tail statistics. The response is decomposed to a statistically stationary part and an intermittent component. The stationary part is quantified using a statistical linearization method while the intermittent part, associated with extreme transient responses, is quantified through i) either a few carefully selected simulations or ii) through the use of effective measures (effective stiffness and damping). The developed approach is able to accurately capture the extreme response statistics orders of magnitude faster compared with direct methods. The scheme is applied to the design and optimization of small attachments that can mitigate and suppress extreme forcing events delivered to a primary structural system. Specifically, we consider the problem of suppression of extreme responses in two prototype ocean engineering systems. First, we consider linear and cubic springs and perform parametric optimization by minimizing the forth-order moments of the response. We then consider a more generic, possibly asymmetric, piece-wise linear spring and optimize its nonlinear characteristics. The resulting asymmetric spring design far outperforms the optimal cubic energy sink and the linear tuned mass dampers.
\end{abstract}

\section{Introduction}

For a plethora of structural systems it is essential to specify their reliability under uncertain environmental loading conditions and most importantly provide design guidelines using knowledge of their response characteristics. This involves accurate estimation of the structural systems probabilistic response. Environmental loads are typically random by nature and are likely to include  intermittently occurring components of an extreme magnitude, representing abnormal environmental events or conditions. Although extreme loadings occur with lower probability than typical conditions, their impact is significant and cannot be neglected since these events determine the systems behavior away from the average operating conditions, which are precisely the conditions that are important to quantify for safe assessment and design. Important examples include mechanical and ocean engineering systems. High speed crafts in rough seas~\cite{Riley2011, Riley2012}, wave impacts on fixed or floating offshore platforms and ship capsize events~\cite{mohamad2016a,belenky07,muller,liu2007,Kreuzer}, vibrations of buildings or bridge structures due to earthquakes or strong wind excitations~\cite{Lin63, Lin98,lin96, gioffre12} are just a few examples where extreme responses occur infrequently but are critical in determining the overall systems reliability. 

Numerous research endeavors have been dedicated on the effective suppression and rapid dissipation
of the  energy associated with extreme impacts on structures. Many of these schemes rely on linear configurations, known as tuned mass damper (TMD) and result in a halving of the resonance frequency. Although the mitigation performance is highly effective when most of the energy is concentrated at the characteristic frequency of the system, their effectiveness drastically drops if there is a mistuning in frequency. Moreover, it is not clear how these configurations perform in the presence of rare impulsive loads. Many of these limitations can be overcome by utilizing small attachments coupled with the primary system through nonlinear springs, also known as nonlinear energy sinks (NES). If carefully chosen these nonlinear attachments can lead to robust, irreversible energy transfer from the primary structure to the attachment and dissipation there~\cite{Vakakis01,
Vakakis08}. The key mechanism behind the efficient energy dissipation in this case is the targeted energy
transfer phenomenon which is an essentially nonlinear mechanism and relies primarily on the energy level of the system, rather then the resonant frequency~\cite{Vakakis03, Kerschen05}. Such configurations have been proven to be successful on the mitigation of deterministic impulsive loads on large structures ~\cite{Shudeifat16, Shudeifat13, Luo14} and their performance has been measured through effective nonlinear measures such as effective damping and stiffness ~\cite{Sapsis12, Quin12}.

Despite their success, nonlinear configurations have been primarily developed for deterministic impulsive loads. To quantify and optimize their performance in the realistic settings mentioned previously it is essential to understand their effects on the statistics of the response and in particular in the heavy tails of the probability distribution function
(PDF). However, quantifying the PDF of nonlinear structures under random forcing containing impulsive type extreme events, poses many challenges for traditional methods. Well established approaches for determining the statistics of nonlinear dynamical systems include the Fokker-Planck equation~\cite{Soong93,sobczyk01}, the joint response-excitation method~\cite{Sapsis08, venturi_sapsis, Joo16, athan_2016_prsa}, Gaussian closure schemes, moment equation or cumulant closure methods~\cite{Beran68, wu84}, the Polynomial-Chaos approach~\cite{Xiu_Karniadakis02}, and stochastic averaging methods~\cite{Zhu88}. For systems associated with heavy tails, however, these methods either cannot capture the statistics of rare and extreme type events due to inherent limitations~\cite{majda_branicki_DCDS} or are far too computationally expensive in practice, even for low-dimensional systems~\cite{Masud_bergman05,di14}. Alternatively, one can study the statistics of the extreme events alone (by ignoring the background `non-extreme' forcing fluctuations) through a Poisson process representation and then analyze the response using the generalized Fokker-Planck or Kolmogorov-Feller equations~\cite{sobczyk01}, which governs the evolution of the corresponding PDF, or by applying the path integral formalism~\cite{koyluoglu95, Iwankiewicz00}, or even through special stochastic averaging techniques~\cite{Zeng11}. While attractive, these ideas lead, in general, to analytical results for a very limited number of special cases.  Besides, it is still an important aspect to account for the background random fluctuations in the forcing term in order to fully characterize the systems overall probabilistic
properties (e.g. this is important in order to fully determine all the moments
of the response). Moreover, even though the background forcing component
does not directly correspond to extreme events, the background term may have
important consequences for the initiation of intermittent type extreme responses~\cite{mohamad2016b}.

In this work we consider the problem of nonlinear structural systems under general time-correlated stochastic forcing that includes extreme, impulsive type random events. We address two important challenges related to this problem. The first is the development of a fast and accurate estimation method for the response statistics, expressed through the PDF, with emphasis on the accurate estimation of the tail form (events far away from the mean). The second is the design and parameter optimization of small attachments that can mitigate or suppress the effects of the extreme forcing events on the system response while they also improve the system behavior during the regular regime. The two problems are connected since extreme event suppression is directly reliant upon a fast and accurate estimation method for the response pdf under different designs or parameters. Indeed, without a fast and reliable method to evaluate response statistics, in particular tail statistics, optimization  cannot be performed because of the inherent computational cost associated with typical quantification methods such as Monte-Carlo. This aspect highlights the practical utility of the proposed fast PDF estimation scheme. We will illustrate the pdf estimation method and shock mitigation design analysis throughout the manuscript with a practical motivating prototype system related to high speed vehicle motion in rough seas, however we emphasize the proposed method broad applicability.

The probabilistic quantification scheme formulated here is based on the most general probabilistic decomposition-synthesis framework~\cite{mohamad2016b,Mohamad15}, that has recently been applied in linear systems subjected to stochastic forcing containing extreme events~\cite{Joo16a} and can be used to efficiently estimate the PDF for the response displacement, velocity, and acceleration. We begin by formulating the response pdf quantification method (developed for linear multi-degree-of-freedom (MDOF) systems in~\cite{Joo16a}) for the case of nonlinear MDOF systems. This is achieved by combining  the probabilistic  decomposition-synthesis framework~\cite{mohamad2016b,Mohamad15} with the statistical linearization method ~\cite{Rob_SPanos03}. The scheme circumvents the rare-event problem and enables rapid design and optimization in the presence of extreme events. We emphasize the statistical accuracy of the derived scheme, which we have validated through extensive comparisons with direct Monte-Carlo simulations. Next, we consider two prototype ocean engineering systems and perform a quantitative comparison of the performance of TMD and NES, evaluating their effectiveness at shock suppression under stochastic excitation containing extreme events. Finally, we perform optimization on a very generic, possibly asymmetric family of piecewise linear springs. Previous endeavors in the context of single-sided vibro-impact NES have shown that asymmetries in the NES can improve the shock mitigation properties (see~\cite{Shudeifat13}). In agreement with these results, our optimization scheme leads to the derivation of a new asymmetric NES which significantly improves the shock mitigation properties of the system in the realistic setting of stochastic excitation.

The paper is structured as follows. In~\cref{sec:proto} we describe the prototype models for high speed craft motion that we utilize throughout the paper as practically relevant example. Next, in~\cref{sec:pds} we provide a brief review of the probabilistic decomposition-synthesis (PDS) framework for the response pdf quantification of a linear single-degree-of-freedom system subject to a random forcing term containing extreme impulse type events. \Cref{sec:quan} describes the proposed general semi-analytical PDF estimation method for nonlinear MDOF structures and also includes a section on quantifying the conditionally rare response  via the effective stiffness and damping framework. In~\cref{sec:optim} we present the mitigation of extreme events analysis on the prototype high speed craft designs for both TMD and cubic NES attachments. Next, in~\cref{sec:designs} we propose a new piecewise linear and asymmetric NES design that we optimize for extreme event mitigation. Finally in~\cref{sec:conc} we offer concluding remarks.

\section{Prototype models for high speed vehicle motion in rough seas}\label{sec:proto}

Here we describe the prototype models that we apply the quantification method for extreme event analysis and optimization. Specifically, we model the motion of a high-speed craft in random seas through two prototype systems: one being a two-degree-of-freedom system consisting of a suspended seat attached to the hull and the second being a three-degree-of-freedom system where the seat is attached to a suspended deck, which is attached on the hull; both prototypes contain a small linear or nonlinear energy sink (NES) vibration absorber.

\subsection{2DOF Suspended seat system}

\begin{figure}[H]
\centering
\includegraphics[scale=1.3]{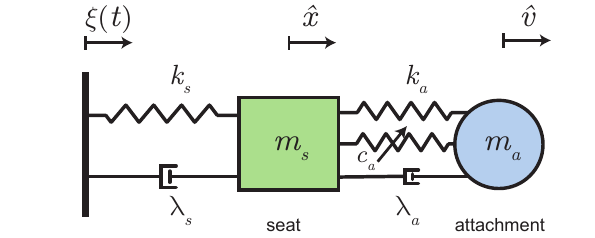}
\caption{[\textbf{Suspended seat}] Mechanical  model for the suspended seat problem with a small attachment (vibration absorber).}
\label{fig:des1}
\end{figure}

In~\cref{fig:des1} we illustrate the first model consisting of a linear primary structure under base excitation that is attached to a small oscillator connected through a nonlinear spring (with cubic nonlineariry). This is a prototype system modeling the suspended
seat of a high speed craft~\cite{Olausson15, Coe09}. The vibration absorber is attached to the seat with the aim to minimize ocean wave impacts on the operator of the vehicle and naturally we require that the attachment mass is much lower than the seat mass (i.e. $m_a < 0.1 m_s$). The equation of motion for this two-degree-of-freedom system is given by:
\begin{align}
m_s\ddot{x} & + \lambda_s \dot{x} + k_s x  +\lambda_a(\dot{x}-\dot{v}) +k_a (x-v) + c_a (x-v)^3 = - m_s \ddot \xi(t),\\
m_a \ddot{v} & + \lambda_a(\dot{v}-\dot{x}) +k_a (v-x) + c_a (v-x)^3 =  - m_a \ddot \xi(t), \nonumber
\label{eq:2dofsys}
\end{align}
where $x,v$ are  the relative displacements of the seat response and attachment response, respectively, with reference to the base motion $\xi(t)$ (that is, $x = \hat{x} - \xi$ and $v = \hat{v} - \xi$).

 \subsection{3DOF Suspended deck-seat system}

\begin{figure}[H]
\centering
\includegraphics[scale=1.3]{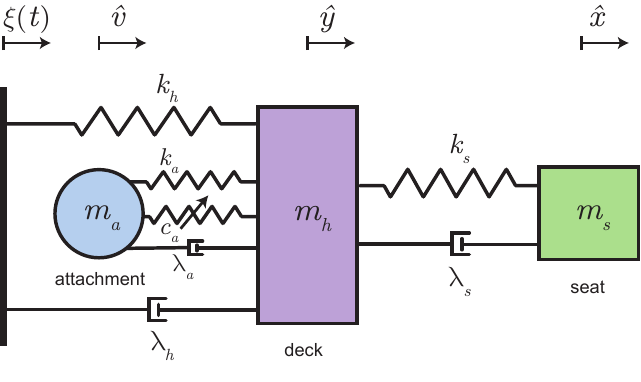}
\caption{[\textbf{Suspended deck-seat}] Mechanical model for the suspended deck-seat problem with a small attachment  (vibration absorber).}
\label{fig:des2}
\end{figure}

The second prototype system is a suspended deck design for a high speed craft~\cite{Townsend12,Ranieri04,KIM97}  and is illustrated in~\cref{fig:des2}. In this case, the vibration absorber is attached to the suspended deck. The attachment mass is comparable to the seat mass and both are considerably smaller than the deck (i.e. $m_a \simeq m_s < 0.1 m_h$). The governing equations for this three-degree-of-freedom system are given by:
\begin{align}
m_h\ddot{y} & + \lambda_h \dot{y} + k_h y  +\lambda_s(\dot{y}-\dot{x}) +k_s (y-x)  +\lambda_a(\dot{y}-\dot{v}) +k_a (y-v) + c_a (y-v)^3 =  - m_h \ddot \xi(t)\\
m_s \ddot{x} & + \lambda_s(\dot{x}-\dot{y}) +k_s (x-y) = - m_s \ddot \xi(t) \nonumber\\
m_a \ddot{v} & + \lambda_a(\dot{v}-\dot{y}) +k_a (v-y) + c_a (v-y)^3 = - m_a \ddot \xi(t)\nonumber,
\label{eq:3dofsys}
\end{align}
where, again, $x,y,v$ are  the relative displacements of the seat response, the deck response and the attachment response, respectively, with reference to the base motion $\xi(t)$.

In both prototypes the aim of the vibration mitigating attachment is to minimize extreme impacts on the seat attachment as this represents an operator on the vehicle. We first examine the case of tuned-mass damper vibration absorber $k_a\neq0, c_a=0$ and the essentially nonlinear energy sink absorber $k_a=0, c_a\neq 0$, that has been studied extensively in the context of shock mitigation \cite{Vakakis08}. In the last section we will examine the performance of an asymmetric, piecewise linear, spring.
 \subsection{The structure of the intermittently extreme stochastic forcing}\label{sec:forc}

Motivated by the ocean engineering systems in~\cref{sec:proto}, we consider
base motion of the form,
\begin{equation}\label{eq:prototype_eg11}
    \ddot{\xi}(t) = \ddot{h}(t) + \sum^{N(t)}_{i=1}  \alpha_i \, \delta(t-\tau_i),
\quad 0 < t \leq T,
\end{equation}
In the expression above, $h(t)$ denotes a zero-mean smooth motion characterized
by a Pierson-Moskowitz spectrum,
\begin{equation}\label{eq:pmspectrum}
    S_{hh}(\omega) = q \frac{1}{\omega^5}\exp\biggl(-\frac{1}{\omega^4}\biggr),
\end{equation}
where $q$ controls the magnitude of the motion. The second term in~\cref{eq:prototype_eg11}
describes rare and extreme impulses in terms of a random impulse train ($\delta(\blank)$
is a unit impulse), occurring due to slamming events. For this component,
$N(t)$ is a Poisson counting process that represents the number of impulses
that arrive in the time interval $0 < t \leq T$, $\alpha$ is the impulse
magnitude, which we assume is normally distributed with mean $\mu_\alpha$
and variance $\sigma_\alpha^2$, and the constant arrival rate is given by
$\nu_r$. We take the impulse magnitude as being $\beta$-times larger than
the standard deviation of the excitation velocity $\dot h(t)$: $\mu_\alpha
= \beta\ \sigma_{\dot{h}}, \text{ with }   \beta\  > 1.$

\section{Review of the probabilistic decomposition-synthesis  (PDS)  method}\label{sec:pds}

We first provide a brief review of the semi-analytical response quantification method for a linear single-degree-of-freedom system~\cite{Joo16a} subjected to stochastic excitation containing rare events. The purpose of this section is to provide a self-contained review of the core ideas, since the scheme for nonlinear structural systems that is described in the following section depends upon these concepts.

Consider the following linear system
\begin{equation}
\label{eq:proto}
    \ddot{x} +\lambda \dot{x} +k x =\ddot{\xi}(t),
\end{equation}
$k$ is the   stiffness, $\lambda$ is the damping, and $\zeta = \lambda/ 2\sqrt{k}$ is the damping ratio. Despite the simplicity of this system, the structure of the statistical response may be significantly complex and posses heavy-tails. 
 
The framework to estimate the response PDF of~\cref{eq:proto} is the probabilistic decomposition-synthesis (PDS) method~\cite{mohamad2016b}. The basic idea is to decouple the rare events regime from the background fluctuations and then  quantify the statistics of the two components separately. The results are then synthesized to obtain the full response PDF by using the total probability law:
 \begin{equation}
    \pdf_x(r) = \pdf_{x_b}(r) (1-\prob_{r}) + \pdf_{x_r}(r)\prob_{r},
\end{equation}
where $\pdf_{x_b}(r)$ is the conditional PDF due to the smooth motion of the base, $\pdf_{x_r}(r)$ is the conditional PDF due to the extreme impacts and $\prob_{r}$ is the overall probability that the system operates in the extreme events regime.

\subsection{Background response PDF}\label{sec:review_bak}

We first obtain the statistical response of the system under the condition that only the background (smooth) forcing component  is acting. We have,
\begin{equation}
    \ddot{x}_b+\lambda \dot{x}_b +k x_b= \ddot{h}(t).
\end{equation}
In this case the analysis is particularly simple since the system is linear and time-invariant and the  response PDF, $\pdf_{x_b}$,  is a zero-mean Gaussian. The spectral density of the response displacement and the variance are given by:
\begin{equation}
    S_{x_bx_b}(\omega) =  \frac{\omega^4S_{hh}(\omega)}{\left(k-\omega^2\right)^2+(\lambda \omega)^2} , \quad \sigma_{x_b}^2 =  \int^\infty_0 S_{x_b x_b}(\omega)\,d\omega.
\end{equation}
The computations for the response velocity and acceleration can be similarly obtained.

\subsection{Numerical histogram for rare events}

The next step is to compute the rare event distribution $ \pdf_{x_r}$ and the rare event probability $\prob_{r}$. Specifically, the rare event distribution can be written as,
\begin{equation}
    \pdf_{x_{r}}(r)  = \int \pdf_{x_{r}\mid \eta}(r\mid n)  \pdf_{\eta}(n)\, dn, 
\end{equation}
where $\pdf_{\eta}(n)$ is the distribution of the impulse magnitude, and $\pdf_{x_{r}\mid \eta}$ is the conditional PDF of the response for an impact of magnitude $\eta$.

It is important to note that once an impulse of magnitude  $\alpha$ hits the system, the momentum of the system right after the impact would be $\dot x_b + \alpha$, since the momentum of the system right before the impact is $\dot x_b$. As these two variables are both Gaussian distributed and independent, their sum is also Gaussian distributed and is given by,
\begin{equation}\label{eq:pdf_eta}
    \eta \equiv \dot x_b + \alpha  \sim \mathcal N(\mu_\alpha,\,  \sigma_{\dot{x}_b}^2 + \sigma_{\alpha}^2).
\end{equation}We estimate the conditional PDF $\pdf_{x_{r}\mid \eta}(r\mid n) $ by  the  numerically computed histogram:
\begin{equation}
    \pdf_{x_r \mid \eta}(r\mid n)  = \hist\bigl\{ x_{r\mid \eta}(t\mid n) \bigr\},  \quad   t \in [0,  \tau_{e}],
\end{equation}
where $\tau_{e}$ is the typical duration of the rare event (see next subsection) and the conditional response $x_{r \mid \eta}$ is given by,
\begin{align}
    x_{r\mid \eta}(t\mid n) =  \frac{n}{2\omega_o}  \biggl(\e^{-(\zeta\omega_n-\omega_o) t}- \e^{-(\zeta\omega_n+\omega_o)  t}\biggr). 
\end{align}
The conditionally extreme event distribution for velocity and acceleration are derived in a similar fashion. 

\subsection{Numerical estimation of the rare event  probability}

In order to compute the histogram of a  rare impulse event, the duration of a rare response needs to be obtained  numerically. We define the typical duration of  a rare response  by
\begin{equation}\label{eq:endtime}
    x_r(\tau_{e})=  \rho_c \, \max\big\{|x_r|\big\},
\end{equation}
where $\rho_c = 0.1$, or in other words, the histogram is taken over the time it takes for the system response to decay to 10\% of its maximum value. The absolute value of the  maximum of the response needs to be estimated numerically. 

Once this rare event duration has been specified, we can also  obtain the probability of a rare event by
\begin{equation}
    \prob_{r}= \nu_\alpha \tau_{e} = \tau_{e}/ T_\alpha.
\end{equation}
Note that the extreme event duration for the displacement $\tau_{e}^{x}$, velocity $\tau_{e}^{\dot x}$, and acceleration $\tau_{e}^{\ddot x}$ are in generally different.

\subsection{Semi-analytical response probability distributions}\label{sec:metho}

With the  description above, we  obtain the  response PDF using the total probability law. The resulting response PDF takes the form,
\begin{align}
    \pdf_{z}(r) = \frac{1 - \nu_\alpha \tau_{e}^{z}}{\sigma_{z_b}\sqrt{2\pi}  }\exp\biggl(-\frac{r^2}{2\sigma_{z_b}^2}\biggr) +  \nu_\alpha \tau_{e}^{z} \int^\infty_0 \hist\bigl \{ z_{r\mid \eta}(t\mid n) \bigr \} \pdf_{\eta}(n)\, dn,
\end{align}
where the argument $z$ is either $x$, $\dot{x}$, or $\ddot{x}$. The validity of this approximation has been thoroughly verified in \cite{Joo16a}.

\section{PDF quantification method for nonlinear MDOF systems}\label{sec:quan}

Here we  formulate the probabilistic-decomposition method for multi-degree-of-freedom, nonlinear mechanical systems. 
There are some important differences with respect to the case of linear systems studied in \cite{Joo16a}. Firstly, for the background component the system nonlinearities can be important and to this end we must utilize an appropriate statistical quantification method.
Here we employ the statistical linearization approach \cite{Rob_SPanos03}.
Secondly, to characterize the statistics in the rare event regime it is even more crucial  to take into account the nonlinear properties of the system, since these control the shock mitigation capabilities of the attachment. 

To achieve this we use two alternative approaches. The first one is based on the direct simulation of the system for a range of initial conditions corresponding to all possible impact magnitudes. The second is based on the notion of effective stiffness and damping~\cite{Sapsis12}, which are measures that characterize the system response under various excitation magnitudes taking into account the presence of the nonlinear attachment.
We provide comparisons with direct Monte-Carlo simulations to demonstrate the accuracy of both approaches. 
We first present the analysis for the background component

\subsection{Quantification of the response pdf for the background component}\label{sec:back}

For the background regime, we must account for nonlinearities and their interaction with the background part of the excitation. We use the statistical linearization method,  since we are only interested in resolving the low-order statistics of the background response of the system (the rare events component defines the tails of the PDF). 

Consider the response of the suspended seat problem,~\cref{eq:2dofsys}, under the excitation term $\ddot h(t)$:
\begin{align}\label{eq:coupled2dofsystem1}
m_s\ddot{x} &\ + \lambda_s \dot{x} + k_s x  +\lambda_a(\dot{x}-\dot{v}) +k_a (x-v) + c_a (x-v)^3 = -m_s\ddot{h}(t),\\
m_a \ddot{v} &\ + \lambda_a(\dot{v}-\dot{x}) +k_a (v-x) + c_a (v-x)^3 = -m_a\ddot{h}(t).
\end{align}
We first multiply the above two equations by $x(s)$, $v(s)$, $h(s)$ at different time instant $s\neq t$, and take ensemble averages to write the resulting equations in terms of covariance functions. 
\begin{align}
&\ m_s C_{xx}''+ \lambda_s C_{xx}' + k_s C_{xx} + \lambda_a\left( C_{xx}'-C_{vx}'\right) + k_a \left(C_{xx}-C_{vx}\right) + c_a \overline{\left(x(t)-v(t)\right)^3x(s)} = -m_s C_{hx}'',\\
&\  m_s C_{xv}''+ \lambda_s C_{xv}' + k_s C_{xv} + \lambda_a\left( C_{xv}'-C_{vv}'\right) + k_a \left(C_{xv}-C_{vv}\right) + c_a \overline{\left(x(t)-v(t)\right)^3v(s)} = -m_s C_{hv}'',\\
&\  m_s C_{xh}''+ \lambda_s C_{xh}' + k_s C_{xh} + \lambda_a\left( C_{xh}'-C_{vh}'\right) + k_a \left(C_{xh}-C_{vh}\right) + c_a \overline{\left(x(t)-v(t)\right)^3h(s)} = -m_s C_{hh}'',\\
&\ m_a C_{vx}'' + \lambda_a\left(C_{vx}'-C_{xx}'\right) +k_a \left(C_{vx}-C_{xx}\right) + c_a \overline{\left(v(t)-x(t)\right)^3x(s)} = -m_a C_{hx}'',\\
&\ m_a C_{vv}'' + \lambda_a\left(C_{vv}'-C_{xv}'\right) +k_a \left(C_{vv}-C_{xv}\right) + c_a \overline{\left(v(t)-x(t)\right)^3v(s)} = -m_a C_{hv}'',\\
&\ m_a C_{vh}'' + \lambda_a\left(C_{vh}'-C_{xh}'\right) +k_a \left(C_{vh}-C_{xh}\right) + c_a \overline{\left(v(t)-x(t)\right)^3h(s)} = -m_a C_{hh}''.
\end{align}
Here $'$ indicates the partial differentiation with respect to the time difference $\tau=t-s$. We then apply Isserlis' theorem based on the Gaussian process approximation for response to express the fourth-order moments in terms of second-order moments \cite{Isserlis18}.
\begin{align}
&\ \overline{\left(x(t)-v(t)\right)^3x(s)} = \left(3 \sigma_x^2 - 6 \sigma_{xv} + 3 \sigma^2_v\right) C_{xx} -  \left(3 \sigma_x^2 - 6 \sigma_{xv} + 3 \sigma^2_v\right) C_{vx},\\
&\ \overline{\left(x(t)-v(t)\right)^3v(s)} = \left(3 \sigma_x^2 - 6 \sigma_{xv} + 3 \sigma^2_v\right) C_{xv} -  \left(3 \sigma_x^2 - 6 \sigma_{xv} + 3 \sigma^2_v\right) C_{vv},\\
&\ \overline{\left(x(t)-v(t)\right)^3h(s)} = \left(3 \sigma_x^2 - 6 \sigma_{xv} + 3 \sigma^2_v\right) C_{xh} -  \left(3 \sigma_x^2 - 6 \sigma_{xv} + 3 \sigma^2_v\right) C_{vh}.
\end{align}
This leads to a set of linear equations in terms of the covariance functions. Thus, the Wiener-Khinchin theorem can be applied to write the equations in terms of the power spectrum, giving
\begin{align}
S_{xx}(\omega; \sigma_x^2, \sigma_{xv}, \sigma_v^2) &= \frac{
\left(m_s+m_a\frac{\mathcal{B}(\omega)}{\mathcal{C}(\omega)}\right)   \left(m_s+m_a\frac{\mathcal{B}(-\omega)}{\mathcal{C}(-\omega)}\right)  
 \omega^4}{\bigl(\mathcal{A}(\omega) -\frac{\mathcal{B}(\omega)^2}{\mathcal{C}(\omega)}\bigr)     \bigl(\mathcal{A}(-\omega) -\frac{\mathcal{B}(-\omega)^2}{\mathcal{C}(-\omega)}\bigr)}S_{hh}(\omega),\label{eq:spectrumrelations3_1}\\
S_{vv}(\omega; \sigma_x^2, \sigma_{xv}, \sigma_v^2) &= \frac{
\left(m_s+m_a\frac{\mathcal{A}(\omega)}{\mathcal{B}(\omega)}\right)  \left(m_s+m_a\frac{\mathcal{A}(-\omega)}{\mathcal{B}(-\omega)}\right)
\omega^4}{\bigl(\frac{\mathcal{A}(\omega)\mathcal{C}(\omega)}{\mathcal{B}(\omega)} - \mathcal{B}(\omega)\bigr)   \bigl(\frac{\mathcal{A}(-\omega)\mathcal{C}(-\omega)}{\mathcal{B}(-\omega)} - \mathcal{B}(-\omega)\bigr)}S_{hh}(\omega),\label{eq:spectrumrelations3_2}\\
S_{xv}(\omega; \sigma_x^2, \sigma_{xv}, \sigma_v^2) &= \frac{
\left(m_s+m_a\frac{\mathcal{B}(\omega)}{\mathcal{C}(\omega)}\right)    \left(m_s+m_a\frac{\mathcal{A}(-\omega)}{\mathcal{B}(-\omega)}\right)  
\omega^4}{\bigl(\mathcal{A}(\omega) -\frac{\mathcal{B}(\omega)^2}{\mathcal{C}(\omega)}\bigr)       \bigl(\frac{\mathcal{A}(-\omega)\mathcal{C}(-\omega)}{\mathcal{B}(-\omega)} - \mathcal{B}(-\omega)\bigr) }S_{hh}(\omega),\label{eq:spectrumrelations3_3}\\
S_{xh}(\omega; \sigma_x^2, \sigma_{xv}, \sigma_v^2) &= \frac{
\left(m_s+m_a\frac{\mathcal{B}(\omega)}{\mathcal{C}(\omega)}\right)  
 \omega^2}{\bigl(\mathcal{A}(\omega) -\frac{\mathcal{B}(\omega)^2}{\mathcal{C}(\omega)}\bigr) }S_{hh}(\omega), \label{eq:spectrumrelations3_4}\\
 S_{vh}(\omega; \sigma_x^2, \sigma_{xv}, \sigma_v^2) &= \frac{
\left(m_s+m_a\frac{\mathcal{A}(\omega)}{\mathcal{B}(\omega)}\right) \omega^2}{\bigl(\frac{\mathcal{A}(\omega)\mathcal{C}(\omega)}{\mathcal{B}(\omega)} - \mathcal{B}(\omega)\bigr) }S_{hh}(\omega),\label{eq:spectrumrelations3_5}
\end{align}
where,
\begin{align}
\mathcal{A}(\omega; \sigma_x^2, \sigma_{xv}, \sigma_v^2) =&  -m_s \omega^2 + (\lambda_s+\lambda_a) (j\omega) + k_s + k_a + c_a(3\sigma_x^2-6\sigma_{xv}+3\sigma_v^2),\label{eq:spectrumrelations22_1}
\\
\mathcal{B}(\omega; \sigma_x^2, \sigma_{xv}, \sigma_v^2) =& \lambda_a (j\omega) + k_a + c_a(3\sigma_x^2-6\sigma_{xv}+3\sigma_v^2), \label{eq:spectrumrelations22_2}
\\
\mathcal{C}(\omega; \sigma_x^2, \sigma_{xv}, \sigma_v^2) =& -m_a \omega^2 + \lambda_a (j\omega) + k_a  + c_a(3\sigma_x^2-6\sigma_{xv}+3\sigma_v^2).
\label{eq:spectrumrelations22_3}
\end{align}
At this point  $\sigma_x^2$, $\sigma_v^2$, and $\sigma_{xv}$ are still unknown, but can be determined by integrating both sides of~\cref{eq:spectrumrelations3_1,eq:spectrumrelations3_2,eq:spectrumrelations3_3} and forming the following system of equations:
\begin{align}
\sigma_x^2 = \int_0^\infty S_{xx}(\omega; \sigma_x^2, \sigma_{xv}, \sigma_v^2)d\omega , \ \ \ \sigma_{xv}= \int_0^\infty S_{xv}(\omega; \sigma_x^2, \sigma_{xv}, \sigma_v^2)d\omega ,\ \ \ \ \sigma_v^2 &= \int_0^\infty S_{vv}(\omega; \sigma_x^2, \sigma_{xv}, \sigma_v^2)d\omega. 
\end{align}
By  solving the above we find  $\sigma_x^2,$ $\sigma_v^2,$ and $\sigma_{xv}$. This procedure  determines the Gaussian PDF approximation for the background regime response. Further details regarding the special case of a linear attachment and the analysis for the suspended deck-seat problem can be found in~\cref{sec:linearization}.\\

\subsection{Quantification of the response pdf for the extreme event component}\label{sec:rare}
We are going to utilize two alternative methods for the quantification of the statistics in the extreme event regime. The first approach is to obtain the conditional statistics based on direct simulations of the system response. The second method is utilizing effective measures~\cite{Sapsis12} that also characterize the system nonlinear response in the presence of attachments.  
\subsubsection{Rare response PDF using direct simulations of the system under impulsive excitation}\label{sec:rarecomp}

To compute the conditionally extreme distribution $p_{x_r}$ and the probability of rare events $\prob_r$ we follow the steps described in ~\cref{alg:rare}, which provides  a high-level description  for a single mode. The procedure is repeated for each degree of freedom of interest (in this case it is more efficient to simply store all the impulse realizations and then run the procedure for each degree of freedom of interest). We emphasize that the numerical simulation of impulse response for nonlinear systems is efficient, since the integrations are necessarily short due the impulsive nature of the forcing and the condition on the rare event end time in~\cref{eq:endtime}. Moreover, throughout these simulations we do not take into account the background excitation since this is negligible compared with the effect of the initial conditions induced by the impact.

\centerline{\begin{minipage}[t]{5in}
    \begin{algorithm}[H]
        \caption{Calculation of $\prob_{r}$ and $\pdf_{x_{r}}(r)  = \int \pdf_{x_{r}\mid \eta}(r\mid n)  \pdf_{\eta}(n)\, dn$.}
        \label{alg:rare}
        \begin{algorithmic}[1]
            \State discretize $\pdf_{\eta}(n)$
            \ForAll{$n$ values over the discretization $\pdf_\eta$}
            \State solve ODE system for $x^n(t)$ under impulse $n$, neglecting $\ddot h$
            \State ${\tau_e}^n \gets  \{ t_e  \mid  \rho_c \max_t{\abs{x^n(t)}} = x^n(t_e) \}$ \Comment{we set $\rho_c = 0.1$}
            \State $ p_{x_r\mid \eta}^{\,n} \gets \hist\bigl\{ x^n(t) \;\big|\; t\in [0, {\tau_e}^n]\bigr\}$
            \EndFor
            \State $ p_{x_r} \gets \int  p_{x_r\mid\eta}^{\,n} \, p_\eta^{\,n} $ 
            \State $\tau_{e} \gets \int \tau_e^{\,n} \,  p_\eta^{\,n} $ 
            \State $\prob_{r} \gets \nu_\alpha \tau_{e}$ 
            \State \textbf{output:} $\prob_{r}, p_{x_r}$
        \end{algorithmic}
    \end{algorithm}
\end{minipage}}

\subsubsection*{Comparison with Monte-Carlo Simulations}\label{sec:mcs}

The full response PDF is composed using the total probability law,
\begin{equation}\label{pdf_rep}
    \pdf_z(r) = \frac{1 - \nu_\alpha \tau_{e, \text{dis}}^z}{\sigma_{z_b}\sqrt{2\pi}  }\exp\biggl(-\frac{r^2}{2\sigma_{z_b}^2}\biggr) +  \nu_\alpha \tau_{e, \text{dis}}^z \int^\infty_0 \hist\bigl \{ z_{r\mid \eta}(t\mid n) \bigr \} \pdf_{\eta}(n)\, dn,
\end{equation}
where $z$ is either the displacement, velocity or acceleration of the seat/attachment response.
We utilize a shifted Pierson-Moskowitz spectrum $S_{hh}(\omega-1)$ for the background forcing term in order to avoid system resonance. Details regarding the Monte-Carlo simulations are provided in~\cref{sec:mcsim}.

\begin{figure}[H]
\centerline{
\begin{minipage}{\hsize}\begin{center}
\includegraphics[width=1\hsize]{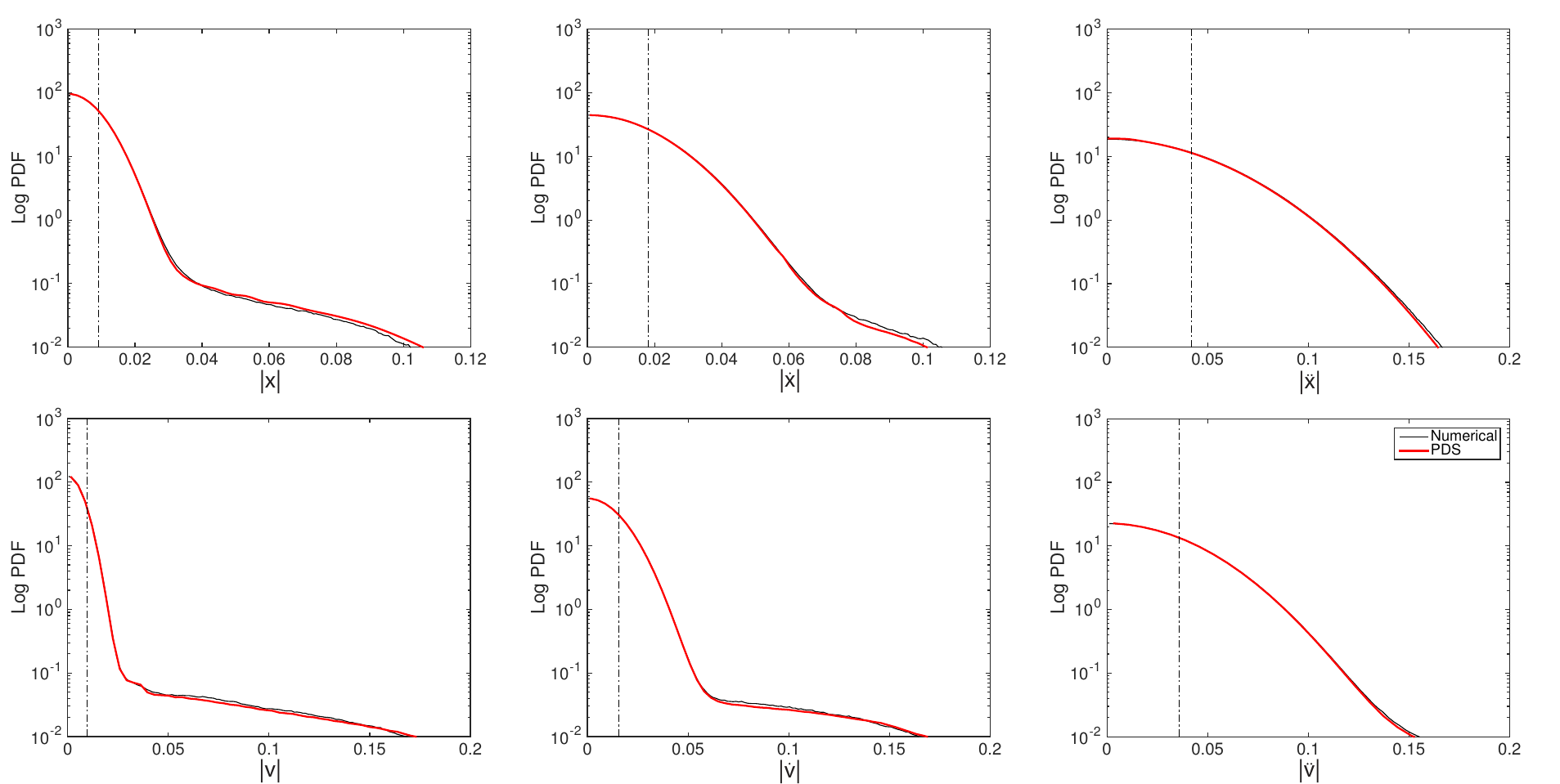}
\end{center}\end{minipage}
}
\caption{Suspended seat  with a NES attached; Comparison between PDS method and Monte-Carlo simulations, with parameters given in~\cref{tab:sdofnes_semi}. Top row: seat response. Bottom row: NES response.}
\label{fig:sdofnes_semi}
\end{figure}

\begin{figure}[H]
\centerline{
\begin{minipage}{\hsize}\begin{center}
\includegraphics[width=1\hsize]{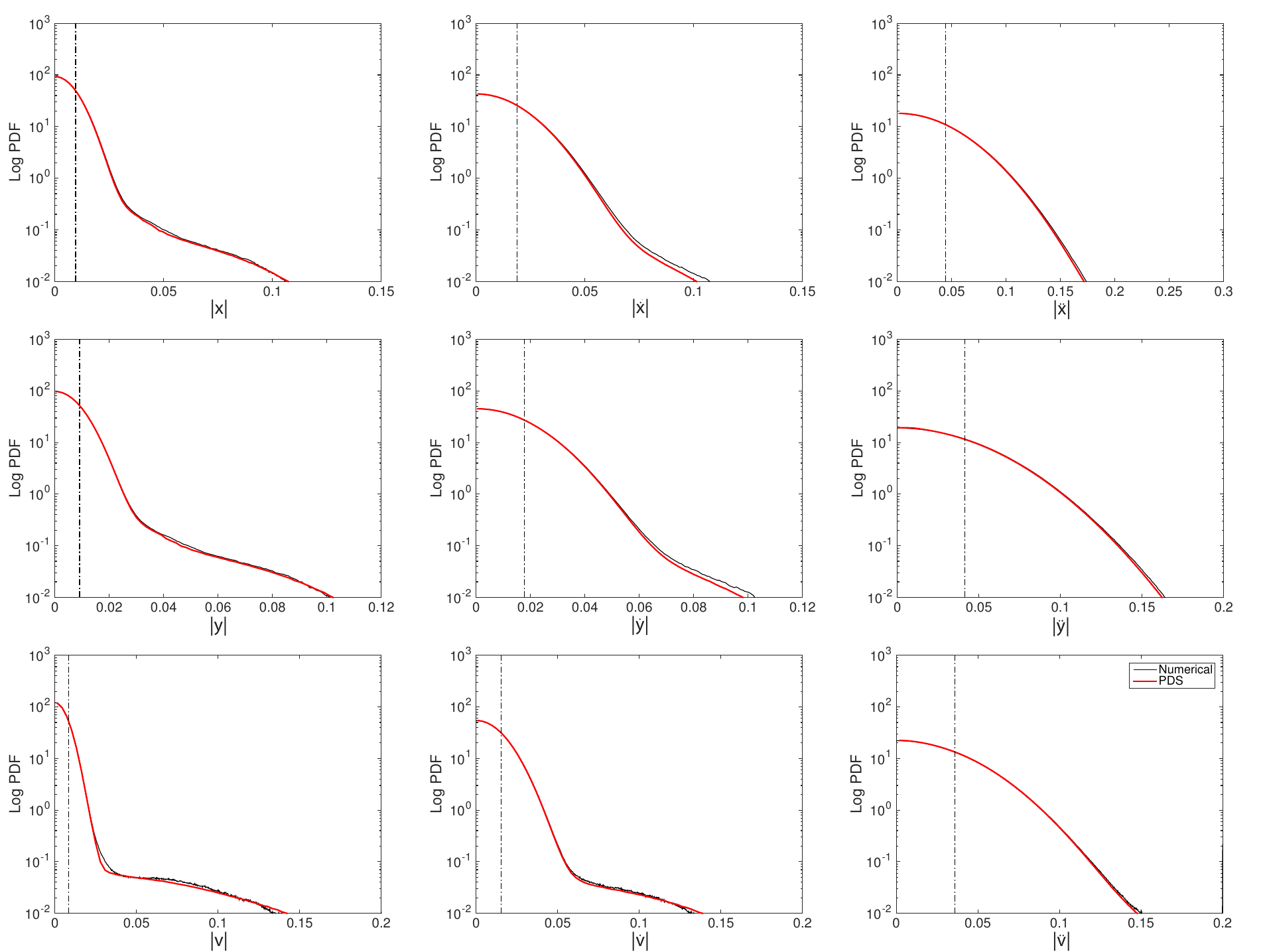}
\end{center}\end{minipage}
}
\caption{Suspended deck-seat with an NES attached; Comparison between the PDS method and Monte-Carlo simulations, with parameters given in~\cref{tab:tdofnes_semi}. Top row: seat response. Middle row: deck response. Bottom row:  NES  response.}
\label{fig:tdofnes_semi}
\end{figure}

In~\cref{fig:sdofnes_semi} we show comparisons for the suspended seat problem with parameters and relevant statistical quantities given in~\cref{tab:sdofnes_semi}. In~\cref{fig:tdofnes_semi} we also show comparisons for the suspended deck-seat problem with parameters and relevant statistical quantities in~\cref{tab:tdofnes_semi}. For both cases the adopted quantification scheme is able to compute the distributions for the quantities of interest extremely fast (less than a minute on a laptop), while the corresponding Monte-Carlo simulations take  order of hours to complete. 

Note that our method is able to capture the complex heavy tail structure many standard deviations away from the mean (dashed vertical line denotes 1 standard deviation). We emphasize that similar accuracy is observed for a variety of system parameters that satisfy the assumptions on the forcing. The close agreement validates that the proposed scheme is applicable and can be accurately used for system optimization and design.  

\begin{table}[H]
\centering
\caption{Parameters and relevant statistical quantities for the suspended seat system. }
{\renewcommand{\arraystretch}{1.2}
\begin{tabular}{l r|lr}
    \toprule
    $m_s$                                & $ 1$     & $m_a$                 & $0.05$                \\
    $\lambda_s$                          & $ 0.01$  & $\lambda_a$           & $0.021$               \\
    $k_s$                                & $ 1$     & $k_a$                 & $0$                   \\
    ---                                  & ---      & $c_a$                 & $ 3.461$              \\
    $T_\alpha$                           & $5000$   & $\sigma_{\eta}$       & $0.0227$              \\
    $\mu_\alpha=7\times\sigma_{\dot{h}}$ & $0.1$    & $q$                   & $1.582\times 10^{-4}$ \\
    $\sigma_\alpha=\sigma_{\dot{h}}$                   & $0.0141$ & $\sigma_{h}$          & $0.0063$              \\
    $\prob_{r}^x$                        & $0.0214$ & $\prob_{r}^v$         & $0.0107$              \\
    $\prob_{r}^{\dot x}$                 & $0.0210$ & $\prob_{r}^{\dot v}$  & $0.0100$              \\
    $\prob_{r}^{\ddot x}$                & $0.0212$ & $\prob_{r}^{\ddot v}$ & $0.0096$              \\
    \bottomrule
\end{tabular}}
\label{tab:sdofnes_semi}
\end{table}

\begin{table}[H]
\centering
\caption{Parameters and relevant statistical quantities for the suspended deck-seat system.}
{\renewcommand{\arraystretch}{1.2}
\begin{tabular}{lr|lr|lr}
    \toprule
    $m_h$                 & $ 1$     & $m_s$                                & $0.05$   & $m_a$                 & $0.05$                \\
    $\lambda_h$           & $ 0.01$  & $\lambda_s$                          & $0.1$    & $\lambda_a$           & $0.035$               \\
    $k_h$                 & $ 1$     & $k_s$                                & $1$      & $k_a$                 & $0$                \\
    ---                   & ---      & ---                                  & ---      & $c_a$                 & $ 5.860$              \\
    $T_\alpha$            & $5000$   & $\mu_\alpha=7\times\sigma_{\dot{h}}$ & $0.1$    & $q$                   & $1.582\times 10^{-4}$ \\
    $\sigma_{\eta}$       & $0.0232$ & $\sigma_\alpha=\sigma_{\dot{h}}$                   & $0.0141$ & $\sigma_{h}$          & $0.0063$              \\
    $\prob_{r}^y$         & $0.0245$ & $\prob_{r}^x$                        & $0.0247$ & $\prob_{r}^v$         & $0.0162$              \\
    $\prob_{r}^{\dot y}$  & $0.0234$ & $\prob_{r}^{\dot x}$                 & $0.0202$ & $\prob_{r}^{\dot v}$  & $0.0161$              \\
    $\prob_{r}^{\ddot y}$ & $0.0238$ & $\prob_{r}^{\ddot x}$                & $0.0081$ & $\prob_{r}^{\ddot v}$ & $0.0146$              \\
\bottomrule
\end{tabular}}
\label{tab:tdofnes_semi}
\end{table}

\subsection{Rare response PDF using effective measures}\label{sec:shocksystems}

Here we describe an alternative technique to quantify the rare event PDF component using the effective stiffness and damping framework described in~\cite{Sapsis12}. These effective measures express any degree-of-freedom of the coupled \emph{nonlinear} system, for a given initial energy level, as an equivalent \emph{linear} single-degree-of-freedom system. Specifically, these effective measures  correspond to the  values of damping and stiffness for a linear system that has (for the same initial conditions) a response that is as close as possible to that of the original system, in the mean square sense.

\begin{figure}[H]
\centerline{\begin{minipage}{\hsize}\begin{center}
\includegraphics[width=0.5\hsize]{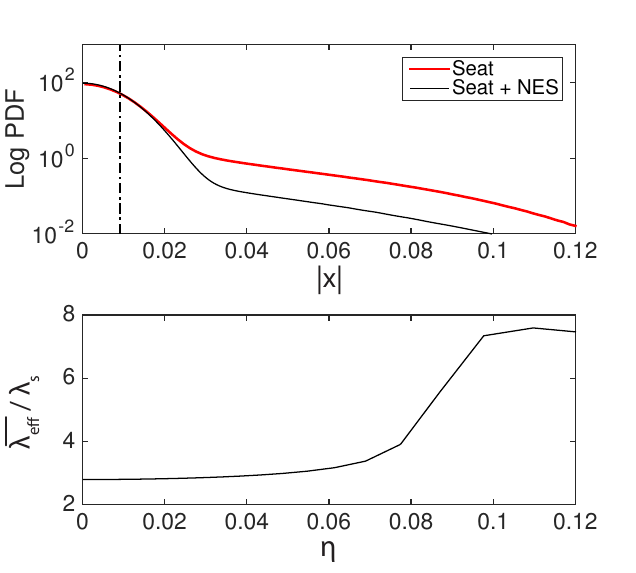}
\end{center}\end{minipage}}
\caption{(Top) Suspended seat problem without and with a NES attached; (Bottom) Normalized weighted-averaged effective damping  $\overline{\lambda}_\text{eff}(n)/\lambda_s$ as a function of  impulse magnitudes $\eta$.}
\label{fig:effsus}
\end{figure}

We focus  on the suspended seat problem to illustrate this strategy. It should be pointed out that the accuracy and applicability of this approach has some limitations:
\begin{itemize}
\item{ The accurate estimation of the PDF requires the knowledge of the effective measures over a sufficiently large range of initial impulses.}
\item{The motion of the system should have an oscillatory character so that it can be captured by effective measures.}
\item{ The statistics of the attachment motion cannot be obtained directly from the effective measures.}

\end{itemize}
To derive the PDF  in the rare event regime  we  reduce the system to an effective linear system for the degree-of-freedome of interest. Consider the suspended seat system under an impulse, 
\begin{equation}\label{eq:2dofrareresp}
\begin{aligned}
m_s\ddot{x} &\ + \lambda_s \dot{x} + k_s x  +\lambda_a(\dot{x}-\dot{v}) +k_a (x-v) + c_a (x-v)^3 =0\\
m_a \ddot{v} &\ + \lambda_a(\dot{v}-\dot{x}) +k_a (v-x) + c_a (v-x)^3 = 0 \end{aligned}
\end{equation}
with   initial conditions, at an arbitrary time say $t_0 = 0$,
\begin{equation}
x = 0,\quad \dot{x}= n ,\quad v = 0 ,\quad  \dot{v} = 0.
\end{equation}
To determine the effective \emph{linear} system for this system, we follow the strategy in~\cite{Sapsis12} and compute the effective stiffness and damping:
\begin{equation}
k_\text{eff}(t;n) =\frac{2\anglebraket{\frac{1}{2}m_s\dot{x}^2}_t}{\anglebraket{ x^2 }_t}, \quad
\lambda_\text{eff}(t;n) =-\frac{\displaystyle 2\frac{d}{dt}\anglebraket{\tfrac{1}{2}m_s\dot{x}^2}_t}{\anglebraket{\dot{x}^2 }_t},
\end{equation}
where $\anglebraket{\blank}$ denotes spline interpolation of the local maxima of the time series. We can then compute the weighted-average effective stiffness and damping:
\begin{equation}
\overline{k}_\text{eff}(n) =\frac{2\int_0^\infty \anglebraket{\tfrac{1}{2}m_s\dot{x}^2}_s \, ds}{ \int_0^\infty \anglebraket{ x^2 }_s \,ds}, \qquad 
\overline{\lambda}_\text{eff}(n) =-2\frac{\int_0^\infty \frac{d}{ds}\anglebraket{\tfrac{1}{2}m_s\dot{x}^2}_s \, ds}{\int_0^\infty \anglebraket{ \dot{x}^2 }_s \,ds}.
\end{equation}
With the weighted-average effective measures we   rewrite the original two-degree-of-freedom system during rare events into an  equivalent linear single-degree-of-freedom system with coefficients that depend on the initial impact  (or the initial energy level of the system):
\begin{equation}\label{eq:effraresys}
\ddot{x} +  \overline{\lambda}_\text{eff}(n) \,  \dot{x} + \overline{k}_\text{eff}(n)\,   x=0
\end{equation}
Using the effective system in~\cref{eq:effraresys} we can obtain the conditionally rare PDF using the analysis for the linear system in~\cref{sec:review_bak}. The damping ratio and natural frequency now become functions of the initial impact,  $n$:
\begin{equation}
\omega_n(n) =  \sqrt{\overline{k}_\text{eff}(n)},\quad
\zeta(n) = \frac{\overline{\lambda}_\text{eff}(n)}{2\sqrt{\overline{k}_\text{eff}(n)}},\quad
\omega_o(n) =  \omega_n(n)\sqrt{\zeta(n)^2-1}.
\end{equation}
Subsequently, the PDF is obtained by taking  a histogram of 
\begin{align}
    x_{r\mid \eta}(t\mid n) =  \frac{n}{2\omega_o(n)}  \biggl(\e^{-(\zeta(n)\omega_n(n)-\omega_o(n))
t}- \e^{-(\zeta(n)\omega_n(n)+\omega_o(n))  t}\biggr). 
\end{align}
In~\cref{fig:effsus} (top) we present the suppression of the probability for large motions of the primary structure due to the presence of the NES (parameters given in~\cref{tab:sdofnes_semi}). This suppression is fully expressed in terms of the effective damping measure shown in the lower plot. Note that the suppression of the tail begins when the effective damping attains values larger than one. 

 We emphasize that in the context of effective measures the motion of the system is  assumed oscillatory. Motions with radically different characteristics will not be captured accurately from the last representation and the resulted histograms will not lead to an accurate representation of the tail. This problem is, in general, circumvented if we employ the first approach for the computation of the conditional PDF during extreme impacts. On the other hand, the advantage of the second approach is that we can interpret the form of the tail in the various regimes with respect to the properties of the effective measures (\cref{fig:effsus}). This link between dynamics (effective measures) and statistics (heavy tail form) is important for the design process of the NES.    

\subsubsection*{Comparison with Monte-Carlo simulations}

\begin{figure}[H]
\centerline{\begin{minipage}{\hsize}\begin{center}
\includegraphics[width=\hsize]{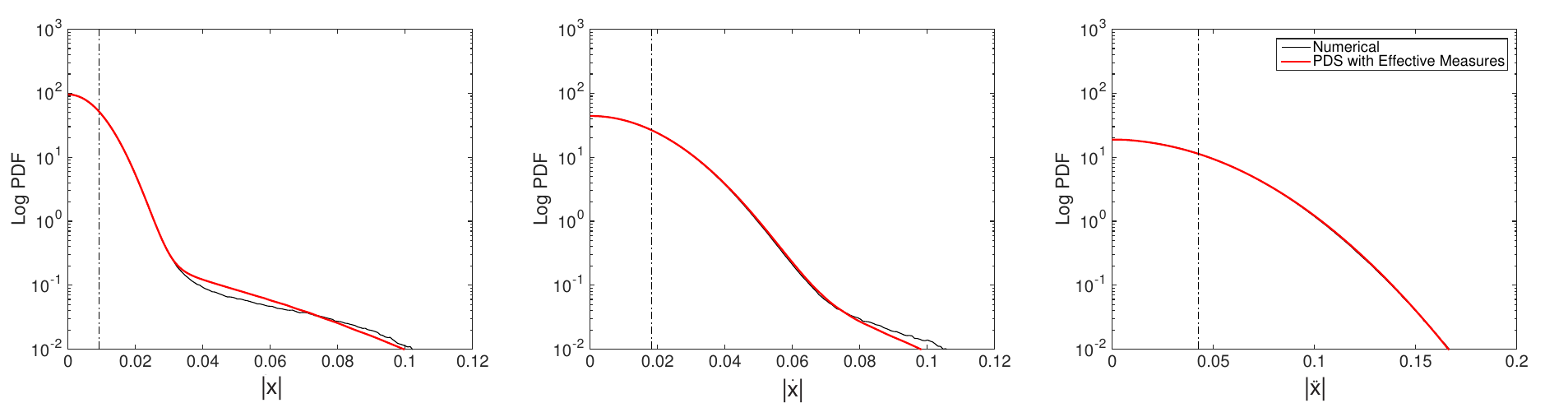}
\end{center}\end{minipage}}
\caption{Suspended seat problem with an NES attached; Comparison between PDS estimate using effective measures and Monte-Carlo simulations. System parameters are given in~\cref{tab:sdofnes_semi}.}
\label{fig:effstiffpdf}
\end{figure}

Here we compare the PDS method combined with the effective measures  with direct Monte-Carlo simulations. In~\cref{fig:effstiffpdf} we show the response PDF for the primary structure for parameters given in~\cref{tab:sdofnes_semi}. Details regarding the Monte-Carlo computations are provided in~\cref{sec:mcs}. We observe that the PDS method utilizing effective measures performs satisfactorily over a wide range similarly with the first general scheme, based on individual trajectories computation. 

% \begin{table}[H]
% \centering
% \caption{Parameters and relevant statistical quantities. }
% {\renewcommand{\arraystretch}{1.2}
% \begin{tabular}{lr|lr}
%     \toprule
%     $m_s$                                & $ 1$     & $m_a$                 & $0.05$                \\
%     $\lambda_s$                          & $ 0.01$  & $\lambda_a$           & $0.021$               \\
%     $k_s$                                & $ 1$     & $k_a$                 & $0$                   \\
%     $T_\alpha$                           & $5000$   & $c_a$                 & $ 3.461$              \\
%     $\mu_\alpha=7\times\sigma_{\dot{h}}$ & $0.1$    & $q$                   & $1.582\times 10^{-4}$ \\
%     $\sigma_{\dot{h}}$                   & $0.0143$ & $\sigma_{h}$          & $0.0063$              \\
%     $\sigma_{\eta}$                      & $0.0230$ & $\prob_{r}^x$         & $0.0162$              \\
%     $\prob_{r}^{\dot x}$                 & $0.0159$ & $\prob_{r}^{\ddot x}$ & $0.0162$              \\
% \bottomrule
% \end{tabular}}
% \label{tab:sdofnes_eff}
% \end{table}

\subsection{Quantification of the absolute response pdf}
The developed  PDF quantification schemes provide statistical description for relative quantities (with respect to the base), that is $x = \hat{x}-\xi$,  $y = \hat{y}-\xi$ and  $v = \hat{v}-\xi$. However, for the prototype systems that we consider we are more interested for the suppression of  absolute quantities, instead of relative ones. As
we illustrate below, the absolute response PDF can be derived from the relative response PDF in a  straightforward manner .  

\subsubsection*{Background component}

For the background regime, we the absolute motion is expressed as:
\begin{equation}
\hat{x}_b = x_b + h.
\end{equation}
As the relative motion and base motion $h(t)$ are both Gaussian distributed (but not independent), their sum is also Gaussian distributed and it is given by,
\begin{equation}
\mathcal{N}(\mu_{\hat{x}}, \sigma_{\hat{x}}^2) = \mathcal{N}(0, \sigma_{{x}}^2+\sigma_{h}^2 + 2 \sigma_{xh}).
\end{equation}
In the previous section we have derived both $\sigma_x^2$ and $\sigma_h^2$, and what remains is the covariance term $\sigma_{xh}$ whose spectral density function is given in \cref{eq:spectrumrelations3_3}. This is given by:
\begin{align}
\sigma_{xh} = \int_0^\infty S_{xh}(\omega; \sigma_x^2, \sigma_{xv}, \sigma_v^2)d\omega.
\end{align}

\subsubsection*{Extreme event component}

For the extreme event component the motion of the motion is assumed very small (compared with the magnitude of the impact), in which case we have:
\begin{equation}
\hat{x}_r = x_r.
\end{equation}
The estimation of the conditional PDF for $x_r$ has already been described in \cref{sec:rare}.

\subsubsection*{Comparison with Monte-Carlo Simulations}

\begin{figure}[H]
\centerline{\begin{minipage}{\hsize}\begin{center}
\includegraphics[width=\hsize]{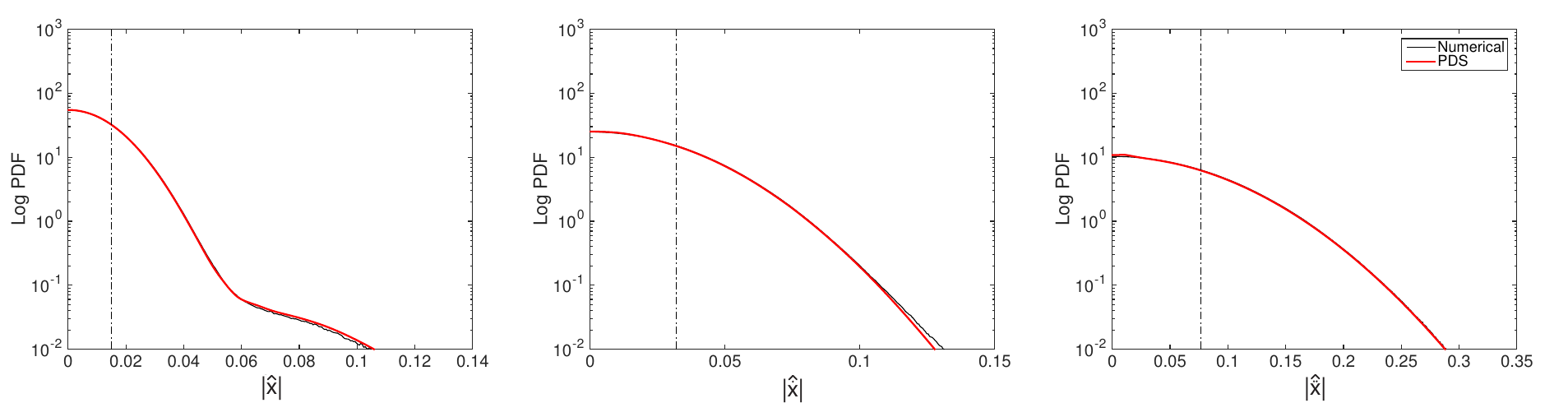}
\end{center}\end{minipage}}
\caption{Suspended seat problem with an NES attached; Comparison between PDS method and Monte-Carlo simulations. System parameters are given in~\cref{tab:sdofnes_semi}.}
\label{fig:absPDF}
\end{figure}

The full absolute response PDF is expressed using eq. (\ref{pdf_rep}),
where $z$ is either relative or absolute displacement, velocity or acceleration of the seat/attachment response. We compare the PDS method with direct Monte-Carlo simulations for the case of absolute motions. In~\cref{fig:absPDF} we show the absolute response PDF for the primary structure for parameters given in~\cref{tab:sdofnes_semi}. Details regarding the Monte-Carlo computations are provided in~\cref{sec:mcs}.

\section{System optimization for extreme event mitigation}\label{sec:optim}

We now consider the problem of optimization in the presence of stochastic excitation containing extreme events.   The developed method provides a rapid and accurate
semi-analytic estimation scheme for the statistical response of the nonlinear structural system. In particular, we can efficiently obtain
the response statistics of the primary structure (the seat) for any given
shock mitigating attachment and accurately capture the heavy-tailed structure
of the distribution.  This allows us to explore rare event mitigation performance
characteristics of different attachment parameters and perform optimization. Such analysis is not practically feasible via
a direct Monte-Carlo approach since a single parameter set takes on the order
of hours to compute the resulting response PDF with converged tail statistics. 

We consider the prototype systems described in~\cref{sec:proto}  with the aim to suppress the large energy delivered to the passenger (i.e. the seat). In all cases we optimize the attachment parameters, while the parameters of the primary structure are assumed to be fixed.

\subsection{Optimization objective}

We adopt the forth-order moment as our measure to reflect the severity of extreme events  on the seat:
\begin{equation}
\overline{\hat{z}^4} =  \int \hat{z}^4 \pdf_{\hat{z}}(r) \, dr,
\end{equation}
where the argument $\hat z$ can be either absolute displacement of the seat or absolute velocity depending on the optimization objective. The goal here is to minimize this measure and   analyze the performance characteristics of the attachment when its  parameters are varied.

We   illustrate the results of the optimization using the following normalized measure:
\begin{equation}
\gamma = {\overline{\hat{z}_a^4}}/{\overline{\hat{z}^4_o}}
\end{equation}
where $\hat{z}_a$ is either $\hat{x}$ or $\hat{\dot{x}}$, and $z_o$ is the corresponding quantity without any attachment. Values of this measure which are less than 1 ($\gamma< 1$) denote   effective  extreme event suppression.

\subsection{Optimization of NES and TMD parameters}

\begin{figure}[H]
\centering
\includegraphics[width=0.8\linewidth]{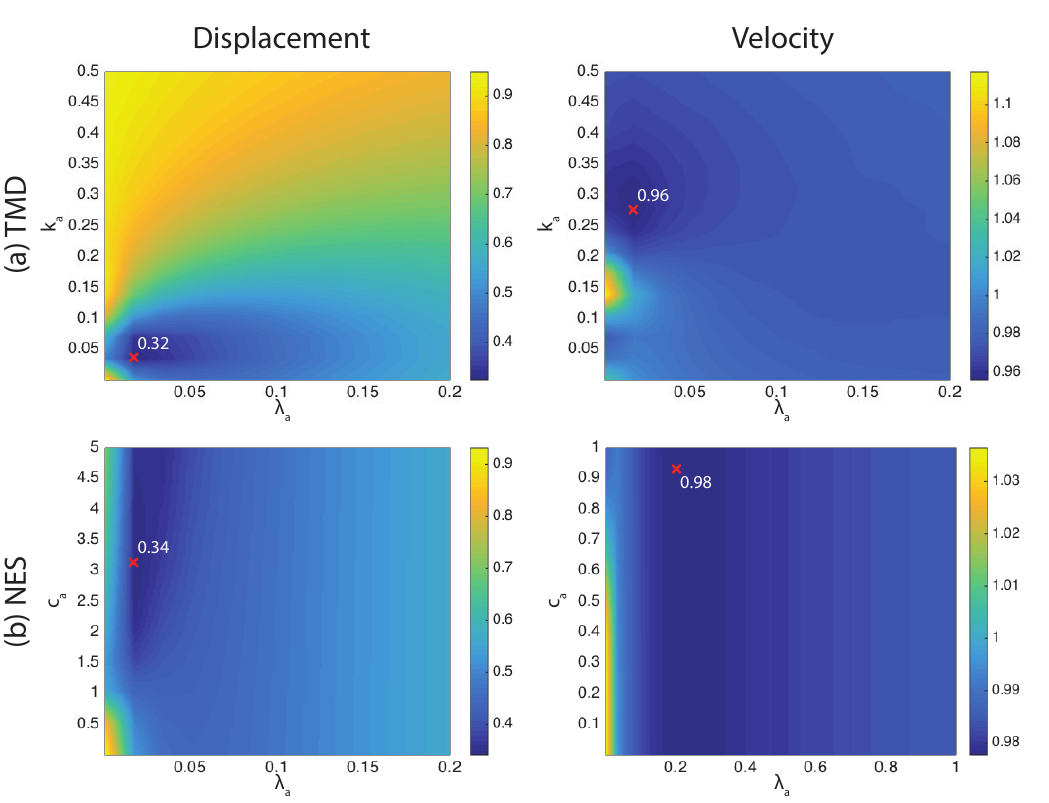}
\caption{[\textbf{Suspended seat}]  The result of the parametric grid search optimization of the suspended seat attached with (a) TMD ($c_a=0$) and (b) NES ($k_a=0$). Optimization has been performed with respect to the stiffness (linear/nonlinear) and damping coefficients of the attachment, and the optimal solutions are marked by a red cross (\textcolor{red}{$\times$}) along with the numeric value of the  optimal measure $\gamma$. Optimization of the response displacement (left subplots) and  velocity (right subplots) are presented. Parameters     without attachment are shown in~\cref{tab:sdof_opt}.}
\label{fig:sdof_opt}
\end{figure}

\begin{figure}[H]
\centerline{
\begin{minipage}{\hsize}\begin{center}
\includegraphics[width=\hsize]{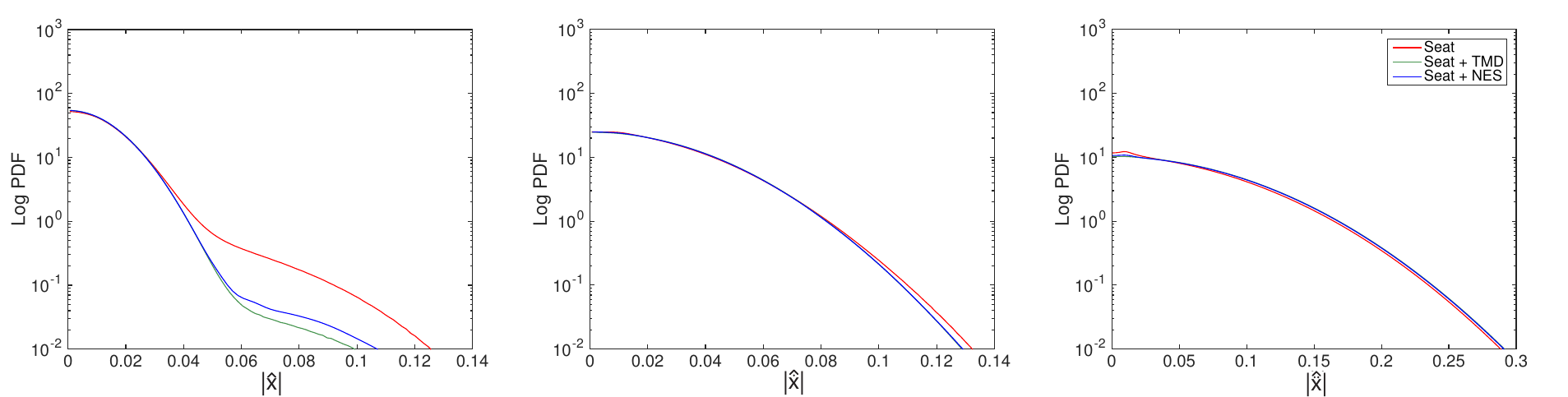}
\end{center}\end{minipage}}
\caption{[\textbf{Suspended seat}] Comparison of the response PDF for optimization of the displacement fourth-order moment. Red curve: without any attachment; Green curve:  TMD  $(\lambda_a=0.018,\, k_a=0.036)$; Blue curve: optimal NES  $(\lambda_a=0.018,\, c_a=3.121)$.}
\label{fig:sdof_opt_pdf}
\end{figure}

\begin{figure}[H]
\centering
\includegraphics[width=0.8\linewidth]{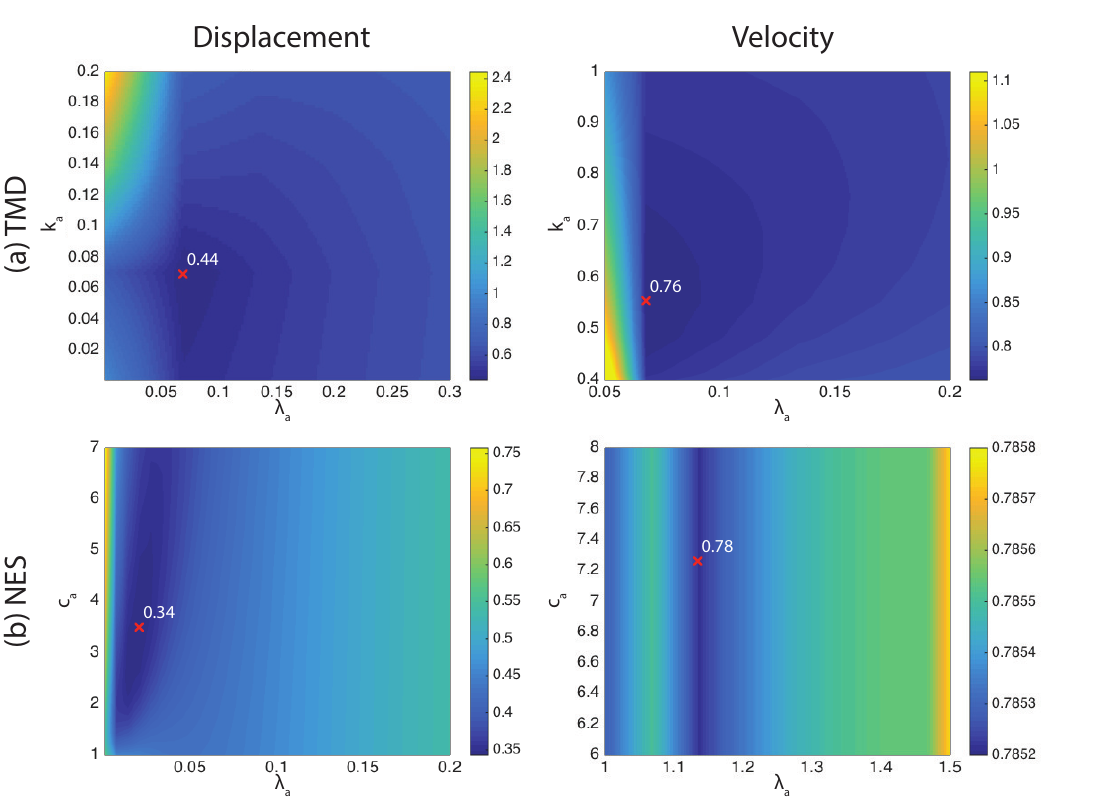}
\caption{[\textbf{Suspended deck-seat}]  The result of parametric grid search optimization of the suspended deck-seat attached with (a) TMD ($c_a=0$) and (b) NES ($k_a=0$). Optimization has been performed with respect to the stiffness (linear/nonlinear) and damping coefficients of the attachment and the optimal solutions are marked by a red cross (\textcolor{red}{$\times$})  along with the numeric value of the  optimal measure $\gamma$. Optimization of  the response displacement (left figures) and velocity (right figures) are presented. Parameters   without attachment are shown in~\cref{tab:tdof_opt}.}
\label{fig:tdof_opt}
\end{figure}

\begin{figure}[H]
\centerline{\begin{minipage}{\hsize}\begin{center}
\includegraphics[width=\hsize]{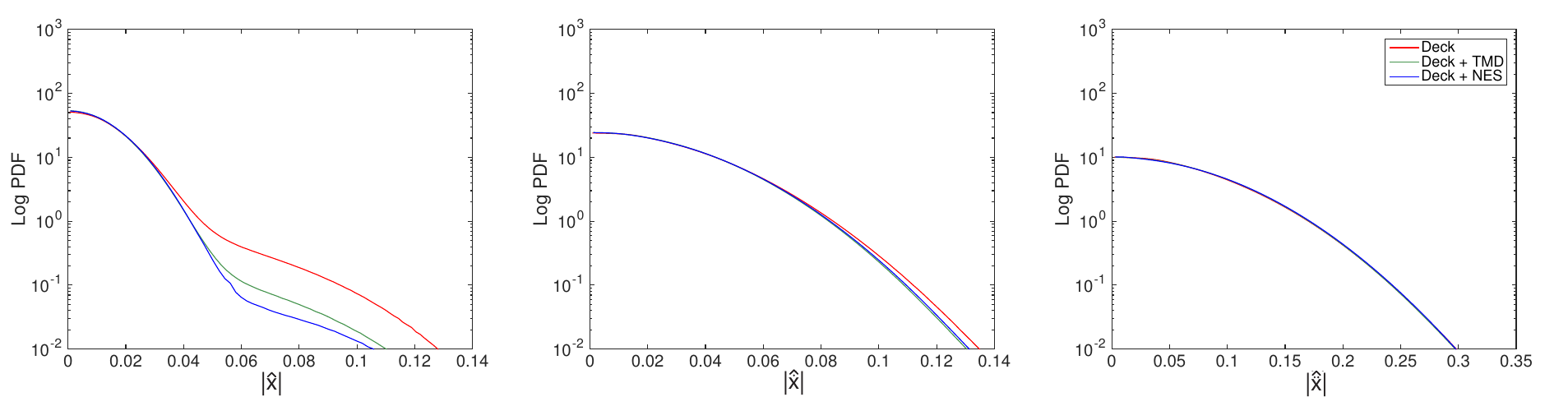}
\end{center}\end{minipage}}
\caption{[\textbf{Suspended deck-seat}] Comparison of the response PDF for optimization
of the displacement fourth-order moment. Red curve: without any attachment; Green curve:  TMD  $(\lambda_a=0.069,\, k_a=0.069)$; Blue curve: optimal NES  $(\lambda_a=0.021,\, c_a=3.484)$.}
\label{fig:tdof_opt_pdf}
\end{figure}

Results are shown for the  suspended seat problem with an attachment mass $m_a = 0.05.$ For a NES attachment ($k_a = 0$) we optimize over $c_a$ and $\lambda_a$, while for a TMD ($c_a = 0$) we vary $k_a$ and $\lambda_a$ (\cref{fig:sdof_opt}). The resulted response PDF  that minimize the    displacement moments  are illustrated in~\cref{fig:sdof_opt_pdf}. The same analysis is performed for the suspended deck-seat problem with the same attachment mass  $m_a = 0.05$ for both systems (\cref{fig:tdof_opt}). The resulted  response PDF  are illustrated in~\cref{fig:tdof_opt_pdf}.

In both cases of systems we observe that the TMD and the optimal cubic NES can improve significantly the behavior of the primary structure in terms of reducing the displacement during impacts, with a reduction of 66-68\% of the fourth-order moment. We also observe that the NES design is more robust to variations in the attachment parameters over the TMD design, which requires more  stringent  attachment parameter values for best performance with respect to $\gamma$. This is in line with the fact that the NES attachment performs better over a broader excitation spectrum than the TMD configuration, which requires carefully tuning. Note that for the case of the deck-seat problem (\cref{fig:tdof_opt}) we can achieve much larger mitigation of the absolute velocity at the order of 32-34\%\ compared with the simpler system of the seat attached to the hull directly (\cref{fig:sdof_opt}), where the suppression is much smaller, 2-4\%.

We performed the grid search for demonstration purposes to illustrate the performance characteristics as the stiffness and damping are varied; clearly, if we are only interested in the optimal attachment the use of an appropriate global optimizer (such a particle swarm optimizer) would be more appropriate. All the results shown where computed using the proposed PDF estimation method. As a further check and validation, we benchmarked the  semi-analytical PDF estimates and compare them  with  Monte-Carlo results for the extremity  measure $\gamma$ over a coarse grid of the attachment parameters.

\begin{table}[H]
\centering
\caption{Suspended seat  system  parameters.}
\begin{center}
{\renewcommand{\arraystretch}{1.2}
\begin{tabular}{lr|lr}
\toprule
$m_s$     &    $ 1$      &      $m_a$      &    $0.05$  \\
$\lambda_s$     &    $ 0.01$      &      $k_s$     &    $ 1$       \\
$T_\alpha$  &    $5000$  &  $-$  & $-$\\
$\mu_\alpha=7\times\sigma_{\dot{h}}$  &    $0.1$            &  $q$  &    $1.582\times 10^{-4}$ \\
$\sigma_\alpha=\sigma_{\dot{h}}$  &    $0.0141$            &  $\sigma_{h}$  &    $0.0063$ \\
\bottomrule
\end{tabular}}
\label{tab:sdof_opt}
\end{center}
\end{table}

\begin{table}[H]
\centering
\caption{Suspended deck-seat  system parameters.}
\begin{center}
\begin{tabular}{lr|lr}
\toprule
$m_h$                                & $ 1$     & $m_s$        & $0.05$                \\
$m_a$                                & $ 0.05$  & $\lambda_h$  & $ 0.01$               \\
$k_h$                                & $ 1$     & $\lambda_s$  & $ 0.1$                \\
$k_s$                                & $ 1$     & $T_\alpha$   & $5000$                \\
$\mu_\alpha=7\times\sigma_{\dot{h}}$ & $0.1$    & $q$          & $1.582\times 10^{-4}$ \\
$\sigma_\alpha=\sigma_{\dot{h}}$                   & $0.0141$ & $\sigma_{h}$ & $0.0063$              \\
\bottomrule
\end{tabular}
\label{tab:tdof_opt}
\end{center}
\end{table}

\section{Design and optimization of a piecewise linear NES}\label{sec:designs}

To further improve the shock mitigation properties of the attachment, we utilize a more generic form of NES consisting of a possibly asymmetric, piecewise linear spring. Similarly with the cubic NES and TMD attachments, we  perform parameter optimization on the NES spring restoring characteristics  and obtain a new optimal design that  outperforms \textcolor{black}{the  TMD and cubic NES} for the considered problems. 

Here, we focus on suppressing large displacements of the seat, although velocity or acceleration would also be appropriate depending on the desired objectives.  The general form of the considered spring consists of a linear regime with slope equal to that of the optimal TMD within a range of 4 standard deviations of the expected seat motion (e.g. when the TMD is employed). For motions (displacements) outside this range the spring has also a linear structure but with different slopes, $\alpha_{-1}$ for negative displacements (beyond 4 standard deviations) and $\alpha_{1}$ for positive displacements (beyond 4 standard deviations).
Therefore, the optimal linear stiffness operates for small to moderate displacement values and outside this regime, when the response is very large, we allow the stiffness characteristics to vary. The objective is to determine the optimal values for the curve in the extreme motion regime with respect to optimization criterion.

Therefore, the analytical form of the piecewise linear spring is given by:
\begin{align}
f (x)= 
\begin{cases}
 \alpha_1 x + \beta_1, \quad & x \geq 4\sigma_\zeta,\\
 k_o x,  \quad &-4\sigma_\zeta\leq x \leq 4\sigma_\zeta,\\
\alpha_{-1} x + \beta_{-1},  \quad  &x\leq -4\sigma_\zeta,
\end{cases}
\end{align}
where, $\sigma_\zeta$ is the standard deviation of the relative displacement $\zeta = x - v$ between the primary structure (the seat) and the attachment for the case of a TMD attachment.
The parameters, $\alpha_1\geq0$ and $\alpha_{-1}\geq0$ define the slopes in the positive and negative extreme response regimes, which we seek to optimize. Moreover, the   values for $\beta_1$ and $\beta_{-1}$ are obtained by enforcing continuity: 
\begin{align}
\beta_1 =&\ 4 (k_o-\alpha_1) \sigma_\zeta,\\
\beta_{-1} =&\ -4 (k_o-\alpha_{-1}) \sigma_\zeta.
\end{align}
The value of the stiffness in the center regime, $k_o$, is chosen using the optimal TMD attachment.

\subsection{Application to the suspended seat and deck-seat problem and comparisons}

We illustrate the  optimization  using the fourth-order moment of the seat response, employing the following measure:
\begin{equation}
\gamma' = {\overline{\hat{z}_n^4}}/{\overline{\hat{z}^4_t}}
\end{equation}
where $\hat{z}_t$ is the system response with the optimal TMD attachment (from the previous parametric grid search optimization) and $\hat{z}_n$ is the response of the system with the piecewise linear NES attachment. Parameters corresponding to values less than 1 ($\gamma'< 1$) denote   additional extreme event suppression, compared with the utilization of optimal TMD. 

\begin{figure}[H]
\centerline{\begin{minipage}{\hsize}\begin{center}
\includegraphics[width=0.7\hsize]{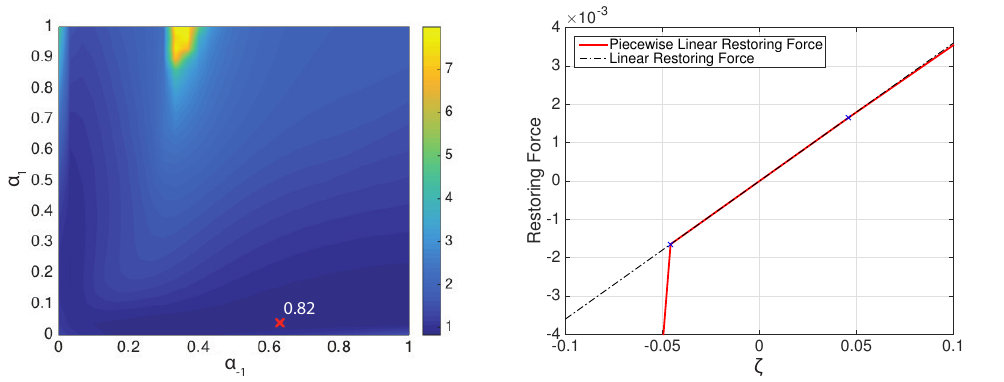}
\end{center}\end{minipage}}
\caption{[\textbf{Suspended seat}] Left: fourth-order  measure $\gamma'$ for the seat absolute displacement as a function of the design variables $\alpha_{-1}$ and $\alpha_{1}$. Right: corresponding  optimal restoring curve ($\alpha_1=0.035,\, \alpha_{-1}=0.634$).}
\label{fig:des_seat_pcolor}
\end{figure}

\begin{figure}[H]
\centerline{\begin{minipage}{\hsize}\begin{center}
\includegraphics[width=\hsize]{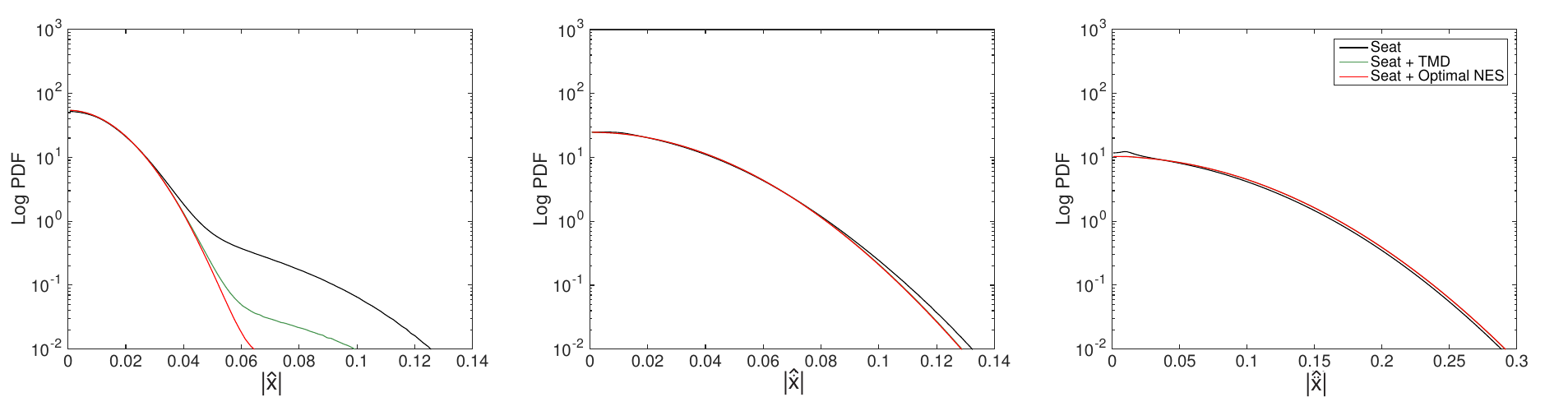}
\end{center}\end{minipage}}
\caption{[\textbf{Suspended seat}] Comparison of the response PDF when the system is tuned for optimal displacement of the seat. Black curve: no attachment. Green curve: optimal TMD design ($\lambda_a=0.018,\, k_a=0.036$). Red curve: proposed optimal piecewise linear NES design.}
\label{fig:des_seat_pdfs}
\end{figure}

\begin{figure}[H]
\centerline{\begin{minipage}{\hsize}\begin{center}
\includegraphics[width=0.75\hsize]{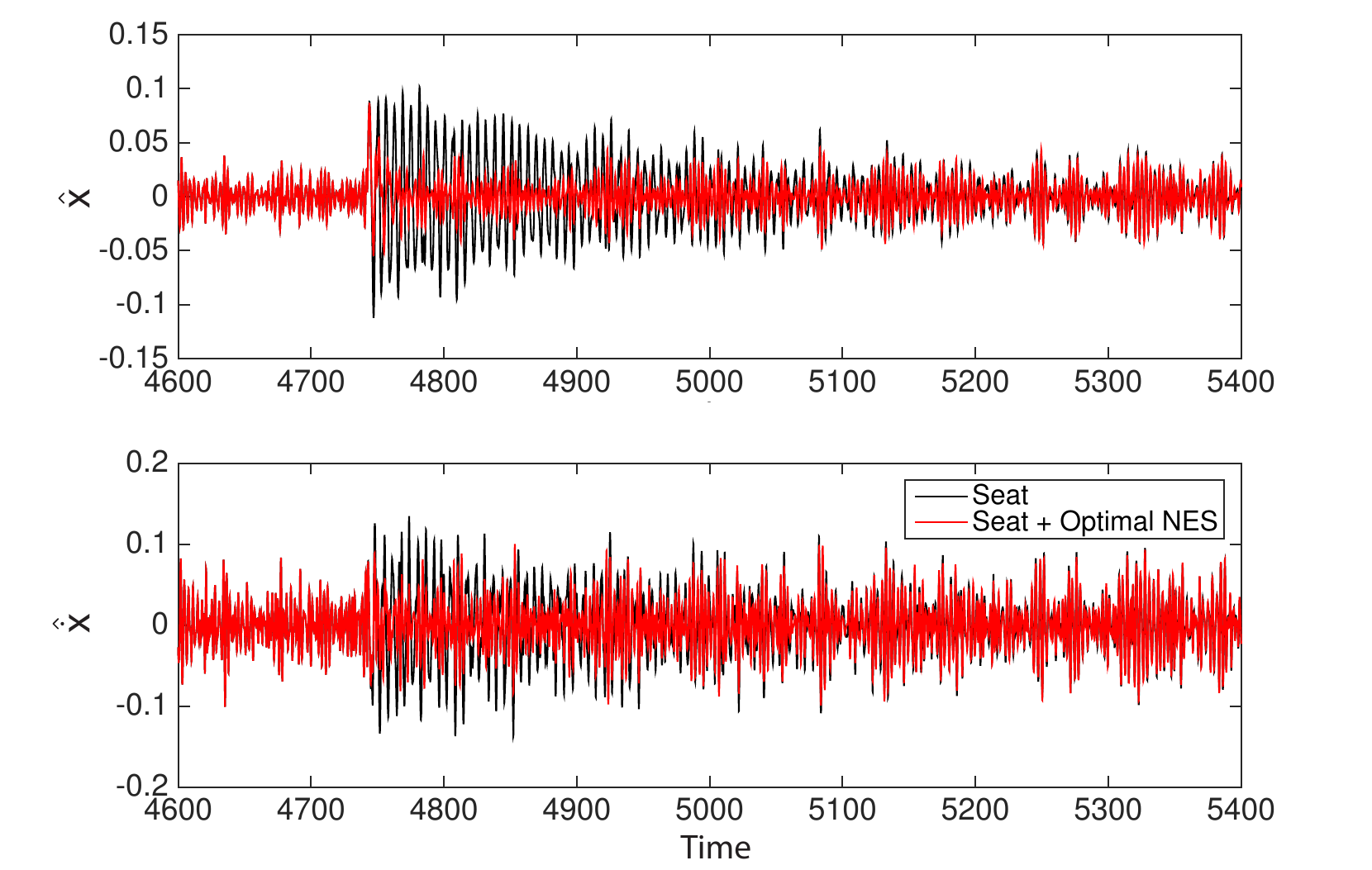}
\end{center}\end{minipage}}
\caption{Representative time series segment for the absolute displacement and velocity
for the  suspended seat problem. Black curve: without
attachment. Red curve: with optimal piece-wise linear NES. This is the result of design
optimization performed in~\cref{fig:des_seat_pcolor}, with 
response PDF shown in~\cref{fig:des_seat_pdfs}.}
\label{fig:timeseries}
\end{figure}

The result of the  optimization for \textit{minimum fourth-order moment for the displacement}, on the suspended seat problem, is shown in~\cref{fig:des_seat_pcolor} while the  corresponding PDF for the displacement, velocity and acceleration are shown in~\cref{fig:des_seat_pdfs}. We note the strongly asymmetric character of the derived piecewise linear spring. This is directly related with the asymmetric character of the impulsive excitation, which is in general positive. 
The performance of the optimized piecewise linear spring is radically improved compared with the optimal cubic NES and TMD as it is shown in the PDF comparisons. Specifically, for rare events (probability of 1\%) we observe a reduction of the motion amplitude by  50\%, while for the velocity the reduction is smaller. A representative time series illustrating the performance of the optimal
 design  for the suspended seat problem is shown in~\cref{fig:timeseries}. 
The PDF for the acceleration for this set of parameters is not changing significantly. Our results are in agreement with previous studies involving
single-sided vibro-impact NES that have been shown to improve  shock mitigation
properties in deterministic setups~\cite{Shudeifat13}.

\begin{figure}[H]
\centerline{\begin{minipage}{\hsize}\begin{center}
\includegraphics[width=0.7\hsize]{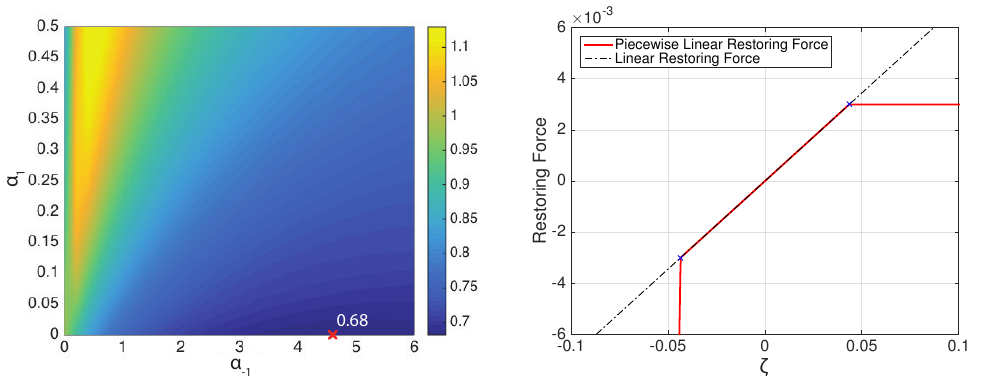}
\end{center}\end{minipage}}
\caption{[\textbf{Suspended deck-seat}] Left: fourth-order  measure $\gamma'$ for the seat absolute displacement as a function of the design variables $\alpha_{-1}$ and $\alpha_{1}$. Right: corresponding  optimal restoring curve ($\alpha_1=0,\, \alpha_{-1}=4.605$).}
\label{fig:des_deck_pcolor}
\end{figure}

\begin{figure}[H]
\centerline{\begin{minipage}{\hsize}\begin{center}
\includegraphics[width=\hsize]{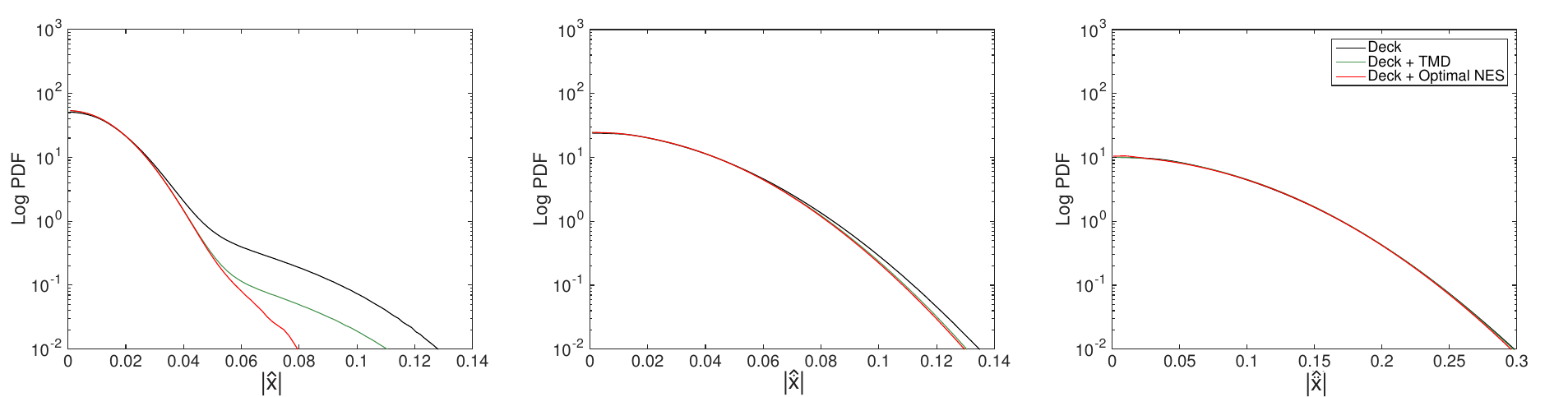}
\end{center}\end{minipage}}
\caption{[\textbf{Suspended deck-seat}] Comparison of the response PDF when the
system is tuned for optimal displacement of the seat. Black curve: no attachment. Green curve: optimal TMD design ($\lambda_a=0.069,\, k_a=0.069$). Red curve: proposed optimal piecewise linear NES design.}
\label{fig:des_deck_pdfs}
\end{figure}

The result of the  optimization for the suspended deck-seat problem is shown in~\cref{fig:des_deck_pcolor} and the  corresponding PDF are shown in~\cref{fig:des_deck_pdfs}. Similarly with the previous problem, the optimization in this case as well leads to a strongly asymmetric piecewise linear spring. The reduction on the amplitude of the displacement during extreme events is radical (with an additional reduction of 32\%) while the corresponding effects for the velocity and acceleration are negligible. This small improvement for the velocity is attributed to the fact that we have focused on minimizing the fourth-order moments for the displacement. 

\section{Summary and conclusions}\label{sec:conc}

We have formulated a parsimonious and accurate quantification method for the heavy-tailed response statistics of nonlinear multi-degree-of-freedom systems under extreme forcing events. The computational core of our approach is the probabilistic decomposition-synthesis method which is formulated for nonlinear MDOF systems under stochastic excitations containing extreme events. Specifically, the excitation is modeled as a superposition of a Poisson distributed impulse train (with extreme magnitude and large inter-arrival times) and a background (smooth) component, modeled by a correlated stochastic excitation with broadband spectral density. This algorithm takes the form of a semi-analytical formula for the response PDF, allowing us to evaluate response statistics (having complex tail structure) on the order of seconds for the nonlinear dynamical structures considered. 

Based on this computational statistical framework, we proceed with the design and optimization of small attachments that can optimally mitigate and suppress the extreme forcing events delivered to the primary system. We performed the suppression of extreme responses on prototype ocean engineering dynamical structures, \emph{the suspended seat} and \emph{the suspended deck-seat} of high speed crafts, via optimal TMD and cubic NES attachments through parametric optimization. As an optimization criterion we selected the forth-order moments of the response displacement, which is a measure of the severity of large deviations from the mean. Quantitative comparisons of TMD and cubic NES were presented, evaluating the effectiveness and robustness in terms of extreme event suppression. We then proposed a new piecewise linear NES with asymmetries, for extreme event mitigation. The optimization of the new design led to a strongly asymmetric spring that far outperforms the optimal cubic NES and TMD for the considered problem.

 We emphasize the statistical accuracy of the PDF estimation schemes, which we demonstrated through comparisons with direct Monte-Carlo simulations. The presented schemes are generic, easy to implement, and can profitably be applied to a variety of different problems in structural engineering where similar characteristics are present, i.e. structures excited by extreme forcing events represented by impulsive-like terms that emerge from an otherwise random excitation background of moderate magnitude.

\section*{Acknowledgments}
We would  like to acknowledge Dr. Timothy Coats, NSWC Carderock, for discussions on the effects of high-speed planing craft wave slams on human performance. T.P.S. has been supported through the ONR grants N00014-14-1-0520 and N00014-15-1-2381 and the AFOSR grant FA9550-16-1-0231. H.K.J. and M.A.M. have been supported through the first and third grants as graduate students. We are also grateful to the Samsung Scholarship Program for support of H.K.J. as well as the MIT Energy Initiative for support under the grant `Nonlinear Energy Harvesting From Broad-Band Vibrational Sources By Mimicking Turbulent Energy Transfer Mechanisms'. 
\appendix

\section{Statistical linearization of the background regime}\label{sec:linearization}

\subsection*{Suspended seat system with a linear attachment}
For the special case of a linear attachment,  $c_a = 0,$   the operators for the suspended seat problem, $\mathcal A,$ $\mathcal B$, and $\mathcal C$ in~\cref{eq:spectrumrelations22_1,eq:spectrumrelations22_2,eq:spectrumrelations22_3} reduce to 
\begin{align}
\mathcal{A}(\omega) &=  -m_s \omega^2 + (\lambda_s+\lambda_a) (j\omega) + k_s + k_a, \\
\mathcal{B}(\omega) &= \lambda_a (j\omega) + k_a, \\
\mathcal{C}(\omega) &= -m_a \omega^2 + \lambda_a (j\omega) + k_a.  
\end{align}
In this case, we can directly integrate~\cref{eq:spectrumrelations3_1,eq:spectrumrelations3_2,eq:spectrumrelations3_3} to obtain  the second order response statistics.

\subsection*{Suspended deck-seat system}
For the suspended deck-seat design the background response is governed by the following system
\begin{align}
m_h\ddot{y} &\ + \lambda_h \dot{y} + k_h y  +\lambda_s(\dot{y}-\dot{x}) +k_s (y-x)  +\lambda_a(\dot{y}-\dot{v}) +k_a (y-v) + c_a (y-v)^3 = -m_h \ddot{h}(t),\\
m_s \ddot{x} &\ + \lambda_s(\dot{x}-\dot{y}) +k_s (x-y) =  -m_s \ddot{h}(t),\\
m_a \ddot{v} &\ + \lambda_a(\dot{v}-\dot{y}) +k_a (v-y) + c_a (v-y)^3 =  -m_a \ddot{h}(t).%\nonumber
\end{align}
As before we first multiply the above two equations by $y(s)$, $x(s)$, $v(s)$, $h(s)$ at different time instant $s\neq t$, and take ensemble averages to write the resulting equations in terms of covariance functions. 
\begin{align}
&\ m_h C_{y\eta}'' +\lambda_h C_{y\eta}' + k_h C_{y\eta}+\lambda_s \left(C_{y\eta}'-C_{x\eta}'\right) +k_s \left(C_{y\eta}-C_{x\eta}\right) \nonumber\\
&\quad\quad\quad\quad\quad\quad\quad\quad +\lambda_a \left(C_{y\eta}'-C_{v\eta}'\right)  +k_a \left(C_{y\eta}-C_{v\eta}\right)  + c_a \overline{\left(y(t)-v(t)\right)^3\eta(s)} = -m_h C_{h\eta}'',\\
&\ m_s C_{x\eta}'' + \lambda_s\left(C_{x\eta}'-C_{y\eta}'\right) + k_s\left(C_{x\eta}-C_{y\eta}\right) = -m_sC_{h\eta}'',\\
&\ m_a C_{v\eta}'' + \lambda_a\left(C_{v\eta}'-C_{y\eta}'\right) + k_a\left(C_{v\eta}-C_{y\eta}\right) +  c_a \overline{\left(v(t)-y(t)\right)^3\eta(s)} =-m_aC_{h\eta}'',
\end{align}
where $\eta$ can be either $y$, $x$, $v$, or $h$, and $'$ indicates the partial differentiation with respect to the time difference $\tau=t-s$. We then apply Isserlis' theorem based on the Gaussian process approximation for response to express the fourth-order moments in terms of second-order moments \cite{Isserlis18}.
\begin{align}
&\ \overline{\left(y(t)-v(t)\right)^3\eta(s)} = \left(3 \sigma_y^2 - 6 \sigma_{yv} + 3 \sigma^2_v\right) C_{y\eta} -  \left(3 \sigma_y^2 - 6 \sigma_{yv} + 3 \sigma^2_v\right) C_{v\eta}.
\end{align}
This leads to a set of linear equations in terms of covariance functions and thus the Wiener-Khinchin theorem can be applied to write the equations in terms of the power spectrum. The spectral equations in this case are given by 
\begin{align}
S_{yy}(\omega; \sigma_y^2, \sigma_{yv}, \sigma_v^2) &= \frac{
\left(m_h+m_a\frac{\mathcal{B}(\omega)}{\mathcal{C}(\omega)}+m_s\frac{\mathcal{D}(\omega)}{\mathcal{E}(\omega)}\right)   \left(m_h+m_a\frac{\mathcal{B}(-\omega)}{\mathcal{C}(-\omega)}+m_s\frac{\mathcal{D}(-\omega)}{\mathcal{E}(-\omega)}\right)
\omega^4\, S_{hh}(\omega)}{\bigl(\mathcal{A}(\omega)-\frac{\mathcal{D}(\omega)^2}{\mathcal{E}(\omega)} - \frac{\mathcal{B}(\omega)^2}{\mathcal{C}(\omega)} \bigr)        \bigl(\mathcal{A}(-\omega)-\frac{\mathcal{D}(-\omega)^2}{\mathcal{E}(-\omega)} - \frac{\mathcal{B}(-\omega)^2}{\mathcal{C}(-\omega)} \bigr)     },\label{eq:spec_3dof_syy}\\
S_{xx}(\omega; \sigma_y^2, \sigma_{yv}, \sigma_v^2) &= 
\frac{\left(m_h+m_s\frac{\mathcal{A}(\omega)}{\mathcal{D}(\omega)}-m_s\frac{\mathcal{B}(\omega)^2}{\mathcal{C}(\omega)\mathcal{D}(\omega)} -m_a\frac{\mathcal{B}(\omega)}{\mathcal{C}(\omega)}  \right)  }
{\bigl(\frac{\mathcal{A}(\omega)\mathcal{E}(\omega)}{\mathcal{D}(\omega)}- \mathcal{D}(\omega)-\frac{\mathcal{B}(\omega)^2\mathcal{E}(\omega)}{\mathcal{D}(\omega)\mathcal{C}(\omega)} \bigr)}\\
& \times \frac{\left(m_h+m_s\frac{\mathcal{A}(-\omega)}{\mathcal{D}(-\omega)}-m_s\frac{\mathcal{B}(-\omega)^2}{\mathcal{C}(-\omega)\mathcal{D}(-\omega)} -m_a\frac{\mathcal{B}(-\omega)}{\mathcal{C}(-\omega)}  \right)   
\omega^4\, S_{hh}(\omega)}
{\bigl(\frac{\mathcal{A}(-\omega)\mathcal{E}(-\omega)}{\mathcal{D}(-\omega)}- \mathcal{D}(-\omega)-\frac{\mathcal{B}(-\omega)^2\mathcal{E}(-\omega)}{\mathcal{D}(-\omega)\mathcal{C}(-\omega)} \bigr)},\label{eq:spec_3dof_sxx}\\
%S_{xx}(\omega; \sigma_y^2, \sigma_{yv}, \sigma_v^2) &= \frac{
%\left(m_h+m_s\frac{\mathcal{A}(\omega)}{\mathcal{D}(\omega)}-m_s\frac{\mathcal{B}(\omega)^2}{\mathcal{C}(\omega)\mathcal{D}(\omega)} -m_a\frac{\mathcal{B}(\omega)}{\mathcal{C}(\omega)}  \right)     \left(m_h+m_s\frac{\mathcal{A}(-\omega)}{\mathcal{D}(-\omega)}-m_s\frac{\mathcal{B}(-\omega)^2}{\mathcal{C}(-\omega)\mathcal{D}(-\omega)} -m_a\frac{\mathcal{B}(-\omega)}{\mathcal{C}(-\omega)}  \right)   
%\omega^4\, S_{hh}(\omega)}{\bigl(\frac{\mathcal{A}(\omega)\mathcal{E}(\omega)}{\mathcal{D}(\omega)}- \mathcal{D}(\omega)-\frac{\mathcal{B}(\omega)^2\mathcal{E}(\omega)}{\mathcal{D}(\omega)\mathcal{C}(\omega)} \bigr)     \bigl(\frac{\mathcal{A}(-\omega)\mathcal{E}(-\omega)}{\mathcal{D}(-\omega)}- \mathcal{D}(-\omega)-\frac{\mathcal{B}(-\omega)^2\mathcal{E}(-\omega)}{\mathcal{D}(-\omega)\mathcal{C}(-\omega)} \bigr)},\label{eq:spec_3dof_sxx}\\
S_{vv}(\omega; \sigma_y^2, \sigma_{yv}, \sigma_v^2) &= 
\frac{\left(m_h+m_a\frac{\mathcal{A}(\omega)}{\mathcal{B}(\omega)}+m_s\frac{\mathcal{D}(\omega)}{\mathcal{E}(\omega)} - m_a\frac{\mathcal{D}(\omega)^2}{\mathcal{E}(\omega)\mathcal{B}(\omega)}\right) }
{\bigl( \frac{\mathcal{A}(\omega)\mathcal{C}(\omega)}{\mathcal{B}(\omega)} -\frac{\mathcal{D}(\omega)^2\mathcal{C}(\omega)}{\mathcal{B}(\omega)\mathcal{E}(\omega)}-\mathcal{B}(\omega)\bigr)}\\
&\times\frac{ \left(m_h+m_a\frac{\mathcal{A}(-\omega)}{\mathcal{B}(-\omega)}+m_s\frac{\mathcal{D}(-\omega)}{\mathcal{E}(-\omega)} - m_a\frac{\mathcal{D}(-\omega)^2}{\mathcal{E}(-\omega)\mathcal{B}(-\omega)}\right) 
\omega^4\, S_{hh}(\omega)}
{ \bigl( \frac{\mathcal{A}(-\omega)\mathcal{C}(-\omega)}{\mathcal{B}(-\omega)} -\frac{\mathcal{D}(-\omega)^2\mathcal{C}(-\omega)}{\mathcal{B}(-\omega)\mathcal{E}(-\omega)}-\mathcal{B}(-\omega)\bigr) }, \label{eq:spec_3dof_svv}\\
%S_{vv}(\omega; \sigma_y^2, \sigma_{yv}, \sigma_v^2) &= \frac{
%\left(m_h+m_a\frac{\mathcal{A}(\omega)}{\mathcal{B}(\omega)}+m_s\frac{\mathcal{D}(\omega)}{\mathcal{E}(\omega)} - m_a\frac{\mathcal{D}(\omega)^2}{\mathcal{E}(\omega)\mathcal{B}(\omega)}\right)    \left(m_h+m_a\frac{\mathcal{A}(-\omega)}{\mathcal{B}(-\omega)}+m_s\frac{\mathcal{D}(-\omega)}{\mathcal{E}(-\omega)} - m_a\frac{\mathcal{D}(-\omega)^2}{\mathcal{E}(-\omega)\mathcal{B}(-\omega)}\right) 
%\omega^4\, S_{hh}(\omega)}{\bigl( \frac{\mathcal{A}(\omega)\mathcal{C}(\omega)}{\mathcal{B}(\omega)} -\frac{\mathcal{D}(\omega)^2\mathcal{C}(\omega)}{\mathcal{B}(\omega)\mathcal{E}(\omega)}-\mathcal{B}(\omega)\bigr)           \bigl( \frac{\mathcal{A}(-\omega)\mathcal{C}(-\omega)}{\mathcal{B}(-\omega)} -\frac{\mathcal{D}(-\omega)^2\mathcal{C}(-\omega)}{\mathcal{B}(-\omega)\mathcal{E}(-\omega)}-\mathcal{B}(-\omega)\bigr)  }, \label{eq:spec_3dof_svv}\\
S_{yv}(\omega; \sigma_y^2, \sigma_{yv}, \sigma_v^2) &= 
\frac{\left(m_h+m_a\frac{\mathcal{B}(\omega)}{\mathcal{C}(\omega)}+m_s\frac{\mathcal{D}(\omega)}{\mathcal{E}(\omega)}\right)}
{ \bigl(\mathcal{A}(\omega)-\frac{\mathcal{D}(\omega)^2}{\mathcal{E}(\omega)} - \frac{\mathcal{B}(\omega)^2}{\mathcal{C}(\omega)} \bigr)}\\
& \times\frac{\left(m_h+m_a\frac{\mathcal{A}(-\omega)}{\mathcal{B}(-\omega)}+m_s\frac{\mathcal{D}(-\omega)}{\mathcal{E}(-\omega)} - m_a\frac{\mathcal{D}(-\omega)^2}{\mathcal{E}(-\omega)\mathcal{B}(-\omega)}\right) 
\omega^4\, S_{hh}(\omega)}
{\bigl( \frac{\mathcal{A}(-\omega)\mathcal{C}(-\omega)}{\mathcal{B}(-\omega)} -\frac{\mathcal{D}(-\omega)^2\mathcal{C}(-\omega)}{\mathcal{B}(-\omega)\mathcal{E}(-\omega)}-\mathcal{B}(-\omega)\bigr) },\label{eq:spec_3dof_syv}\\
%S_{yv}(\omega; \sigma_y^2, \sigma_{yv}, \sigma_v^2) &= \frac{
%\left(m_h+m_a\frac{\mathcal{B}(\omega)}{\mathcal{C}(\omega)}+m_s\frac{\mathcal{D}(\omega)}{\mathcal{E}(\omega)}\right)    \left(m_h+m_a\frac{\mathcal{A}(-\omega)}{\mathcal{B}(-\omega)}+m_s\frac{\mathcal{D}(-\omega)}{\mathcal{E}(-\omega)} - m_a\frac{\mathcal{D}(-\omega)^2}{\mathcal{E}(-\omega)\mathcal{B}(-\omega)}\right) 
%\omega^4\, S_{hh}(\omega)}{ \bigl(\mathcal{A}(\omega)-\frac{\mathcal{D}(\omega)^2}{\mathcal{E}(\omega)} - \frac{\mathcal{B}(\omega)^2}{\mathcal{C}(\omega)} \bigr)       \bigl( \frac{\mathcal{A}(-\omega)\mathcal{C}(-\omega)}{\mathcal{B}(-\omega)} -\frac{\mathcal{D}(-\omega)^2\mathcal{C}(-\omega)}{\mathcal{B}(-\omega)\mathcal{E}(-\omega)}-\mathcal{B}(-\omega)\bigr)   },\label{eq:spec_3dof_syv}
S_{yh}(\omega; \sigma_y^2, \sigma_{yv}, \sigma_v^2) &= \frac{
\left(m_h+m_a\frac{\mathcal{B}(\omega)}{\mathcal{C}(\omega)}+m_s\frac{\mathcal{D}(\omega)}{\mathcal{E}(\omega)}\right)   
\omega^2\, S_{hh}(\omega)}{\bigl(\mathcal{A}(\omega)-\frac{\mathcal{D}(\omega)^2}{\mathcal{E}(\omega)} - \frac{\mathcal{B}(\omega)^2}{\mathcal{C}(\omega)} \bigr)        },\label{eq:spec_3dof_syh}\\
S_{xh}(\omega; \sigma_y^2, \sigma_{yv}, \sigma_v^2) &= 
\frac{\left(m_h+m_s\frac{\mathcal{A}(\omega)}{\mathcal{D}(\omega)}-m_s\frac{\mathcal{B}(\omega)^2}{\mathcal{C}(\omega)\mathcal{D}(\omega)} -m_a\frac{\mathcal{B}(\omega)}{\mathcal{C}(\omega)}  \right)  \omega^2\, S_{hh}(\omega)}
{\bigl(\frac{\mathcal{A}(\omega)\mathcal{E}(\omega)}{\mathcal{D}(\omega)}- \mathcal{D}(\omega)-\frac{\mathcal{B}(\omega)^2\mathcal{E}(\omega)}{\mathcal{D}(\omega)\mathcal{C}(\omega)} \bigr)},
\label{eq:spec_3dof_sxh}\\
S_{vh}(\omega; \sigma_y^2, \sigma_{yv}, \sigma_v^2) &= 
\frac{\left(m_h+m_a\frac{\mathcal{A}(\omega)}{\mathcal{B}(\omega)}+m_s\frac{\mathcal{D}(\omega)}{\mathcal{E}(\omega)} - m_a\frac{\mathcal{D}(\omega)^2}{\mathcal{E}(\omega)\mathcal{B}(\omega)}\right) \omega^2\, S_{hh}(\omega)}
{\bigl( \frac{\mathcal{A}(\omega)\mathcal{C}(\omega)}{\mathcal{B}(\omega)} -\frac{\mathcal{D}(\omega)^2\mathcal{C}(\omega)}{\mathcal{B}(\omega)\mathcal{E}(\omega)}-\mathcal{B}(\omega)\bigr)},\label{eq:spec_3dof_svh}
\end{align}
where
\begin{align}
\mathcal{A}(\omega; \sigma_y^2, \sigma_{yv}, \sigma_v^2)&= -m_h \omega^2 + (\lambda_h+\lambda_s+\lambda_a) (j\omega) + k_h + k_s + k_a + c_a (3\sigma_y^2 - 6\sigma_{yv} + 3\sigma_v^2), \label{eq:susd_1} \\
\mathcal{B}(\omega; \sigma_y^2, \sigma_{yv}, \sigma_v^2) &= \lambda_a (j\omega) + k_a + c_a (3\sigma_y^2 - 6\sigma_{yv} + 3\sigma_v^2),  \label{eq:susd_2} \\
\mathcal{C}(\omega; \sigma_y^2, \sigma_{yv}, \sigma_v^2) &= -m_a \omega^2 + \lambda_a (j\omega) + k_a + c_a (3\sigma_y^2 - 6\sigma_{yv} + 3\sigma_v^2),  \\
\mathcal{D}(\omega) &= \lambda_s (j\omega) + k_s, \\
\mathcal{E}(\omega) &= -m_s \omega^2 + \lambda_a (j\omega) + k_s. 
\label{eq:susd_3} 
\end{align}
Now  $\sigma_y^2$, $\sigma_v^2$, and $\sigma_{yv}$ are still unknown, but can be determined by integrating both sides of \cref{eq:spec_3dof_syy,eq:spec_3dof_sxx,eq:spec_3dof_svv,eq:spec_3dof_syv} and forming the following system of equations,
\begin{align}
\sigma_y^2&=\int_0^\infty S_{xx}(\omega; \sigma_y^2, \sigma_{yv}, \sigma_v^2)d\omega, \\   
 \sigma_{yv} &= \int_0^\infty S_{yv}(\omega; \sigma_y^2, \sigma_{yv}, \sigma_v^2)d\omega, \\
 \sigma_v^2 &= \int_0^\infty S_{vv}(\omega; \sigma_y^2, \sigma_{yv}, \sigma_v^2)d\omega,
\end{align}
from which we obtain $\sigma_y^2, \sigma_{yv}, \sigma_v^2 $.

\subsection*{Suspended deck-saet system with a linear attachment} If the  attachment is linear  $c_a = 0$, $\mathcal A,$ $\mathcal B$, and $\mathcal C$ in~\cref{eq:susd_3,eq:susd_2,eq:susd_1} reduce to 
\begin{align}
\mathcal{A}(\omega) &=  -m_h \omega^2 + (\lambda_h+\lambda_s+\lambda_a) (j\omega) + k_h + k_s + k_a, \\
\mathcal{B}(\omega) &= \lambda_a (j\omega) + k_a, \\
\mathcal{C}(\omega) &= -m_a \omega^2 + \lambda_a (j\omega) + k_a , 
\end{align}
which can be directly integrated to obtain  the second order response statistics.

\section{Monte-Carlo simulations}\label{sec:mcsim}

For the Monte-Carlo simulations the excitation time series is generated by superimposing the background and rare event components. The background excitation, described by a stationary stochastic process with a Pierson-Moskowitz spectrum (\cref{eq:pmspectrum}), is simulated through  a superposition of cosines over a range of frequencies with corresponding amplitudes  and uniformly distributed random phases. The intermittent  component  is the random impulse train, and each impact is  introduced as a  velocity jump at the point of the impulse. For each of the comparisons performed in this work we generated $10$ realizations of the excitation time series, each with a train of $100$ impulses. Once each ensemble  for the excitation is  computed, the governing ordinary differential equations  are solved using a 4th/5th order Runge-Kutta method (we carefully  account for the modifications in the momentum that an impulse imparts by integrating up to each impulse time and modifying the initial conditions that the impulse imparts before integrating the system to the next impulse time). We verified that this number of ensembles and their durations leads to converged  response statistics for the displacement, velocity, and acceleration.

\printbibliography

\end{document}